\documentclass[conf]{new-aiaa}
\usepackage[utf8]{inputenc}

\usepackage{geometry}
\usepackage{graphicx}
\usepackage{amsmath}
\usepackage[version=4]{mhchem}
\usepackage{siunitx}
\usepackage{longtable,tabularx}
\usepackage{color}

\usepackage{float}
\usepackage{pgfplots}
\usepackage{pgfplotstable}
\usepackage{tikz}
\usetikzlibrary{arrows.meta}
\usetikzlibrary{positioning}
\usepackage{ar}
\usepackage[numbers]{natbib}
\usepackage{hyperref}
\usepackage[]{caption}

\usepackage{tabularx}
\usepackage{booktabs}

\usepackage{xcolor}
\usepackage{subfig}
\usepackage{multicol}

\newcommand{\myvec}[1]{\boldsymbol{#1}}

\def \vx{\boldsymbol{x}}

\def \vv{\myvec{V}}

\newcolumntype{M}[1]{>{\centering\arraybackslash}m{#1}}

\title{Machine Learning to Predict Aerodynamic Stall}

\author[1]{Ettore Saetta\footnote{PhD student, Department of Industrial Engineering, University of Naples Federico II.}}
\author[1]{Renato Tognaccini\footnote{Associate Professor, Department of Industrial Engineering, University of Naples Federico II.}}
\author[2]{Gianluca Iaccarino\footnote{Professor, Department of Mechanical Engineering, Stanford University.}}
\affil[1]{University of Naples Federico II, Naples 80125, Italy}
\affil[2]{Stanford University, Stanford, CA 94305, USA}

\begin{document}

\maketitle

\begin{abstract}
 A {convolutional autoencoder} is trained using a database of  airfoil aerodynamic simulations and assessed in terms of overall accuracy and interpretability.
The goal is to  predict  the stall and to investigate the ability of the autoencoder to distinguish between the linear and non-linear response of the airfoil pressure distribution to changes in the angle of attack.
After a sensitivity analysis on the  learning  infrastructure, we investigate the latent space identified by the autoencoder targeting extreme compression rates, i.e. very low-dimensional reconstructions. We also propose a strategy to use the decoder to generate new \textit{synthetic} airfoil geometries and aerodynamic solutions by interpolation and extrapolation in the latent representation learned by the autoencoder.
\end{abstract}


\section{Introduction}


Experimental and computational techniques to investigate aerodynamic performance  have matured in the last decades providing increasingly  large amounts of quantitative information. These extensive datasets and the recent success of machine learning (ML) techniques in a variety of applications have spurred interest in developing data-driven techniques for fluid mechanics. 
Artificial neural networks and other supervised ML techniques algorithms have demonstrated promising potential to improve turbulence models \cite{Duraisamy2019, Duraisamy2015, Tracey2015, Milano2002}, to accelerate shape optimization \cite{Yan2019, Li2020} and to investigate flow control strategies \cite{Noack2018, Verma2018}.

In this work we investigate  the ability of an unsupervised ML strategy, \textit{convolutional autoencoders},  to predict aerodynamic characteristics and specifically the stall of classic wing cross-sections.

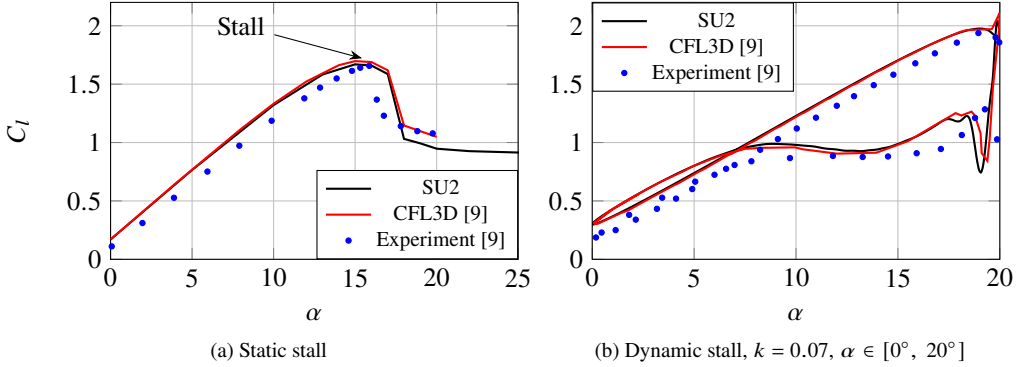
\begin{figure}[H]
    \setcounter{subfigure}{0}
    \centering
    \subfloat[Static stall]{\begin{tikzpicture}
\begin{axis}[height=5cm, width=7cm, grid, xmin=0, xmax = 25,ymin=0,ymax=2.2,xlabel=$\alpha$, ylabel=$C_l$,
legend style={anchor= south east, at={(1.0,0.0)},nodes={scale=0.8}}]

\node at (80,200) (nodeA) {Stall};
\node at (160,170) (nodeB) {};
    
\draw [-{Stealth}] (nodeA) -- (nodeB);

\addplot[thick,black]
table [x expr=\thisrowno{0}, y expr=\thisrowno{1}] {Plots/cfdM0p3_static_dim.dat};

\addplot[thick,red]
table [x expr=\thisrowno{0}, y expr=\thisrowno{1}] {Plots/cfd_nasa_M0p3_static.dat};

\addplot[only marks,blue,mark size=1pt]
table [x expr=\thisrowno{0}, y expr=\thisrowno{1}] {Plots/expM0p3_static.dat};

\legend{SU2, CFL3D \cite{matalanis2016}, Experiment \cite{matalanis2016}}

\end{axis}
\end{tikzpicture}\label{subfig.static_stall}}
\subfloat[{Dynamic stall, $k=0.07$, $\alpha\in\left[0^\circ,\ 20^\circ\right]$}]{\begin{tikzpicture}
\begin{axis}[height=5cm, width=7cm, grid, xmin=0, xmax = 20,ymin=0,ymax=2.2,xlabel=$\alpha$,
legend style={anchor= north west, at={(0.0,1.0)},nodes={scale=0.8}}]

\addplot[thick,black,each nth point={10}]
table [x expr=\thisrowno{0}, y expr=\thisrowno{1}] {Plots/unst_cfdM0p3_correct.dat};

\addplot[thick,red]
table [x expr=\thisrowno{0}, y expr=\thisrowno{1}] {Plots/unst_cfdM0p3_nasa.dat};

\addplot[only marks,blue,mark size=1pt,each nth point={3}]
table [x expr=\thisrowno{0}, y expr=\thisrowno{1}] {Plots/unst_expM0p3_nasa.dat};

\legend{SU2, CFL3D \cite{matalanis2016}, Experiment \cite{matalanis2016}}

\end{axis}
\end{tikzpicture}\label{subfig.dynamic_stall}}
    
\caption{VR12 airfoil lift curves, $M_\infty=0.3$, $Re_\infty=2.6\times10^6$. Comparison between CFD simulation present authors obtained using SU2, CFD results by Matalanis et al. using CFL3D and experimental data \cite{matalanis2016}.}
\label{fig.lift_curves}
\end{figure}

 Airfoil stall is a strongly non-linear phenomenon corresponding to the loss of lift force and a primary design consideration for airplanes and rotor-crafts. 
Fig. \ref{fig.lift_curves} illustrates the lift curves (variation of lift coefficient $C_l$ as function of the angle of attack $\alpha$) for the Boeing VR12 airfoil in steady (\ref{subfig.static_stall}) and unsteady (\ref{subfig.dynamic_stall}) regimes, experiment by Matalanis et al. \cite{matalanis2016}.
The figure also shows the capabilities of CFD simulations as obtained by present authors (SU2 RANS and URANS solver \cite{Economon2016SU2}) and by Matalanis et al. (CFL3D) in predicting these phenomena.

The stall condition corresponds to the maximum lift. For fixed wing aircraft at low angle of attack, classical aerodynamic theories predict a lift coefficient ($C_l$) that varies linearly with $\alpha$; on the other hand for $\alpha>10^\circ$ a strong non linear behavior is observed with the lift decreasing after the stall angle due mainly to the presence of  turbulent separated flow on the upper surface of the airfoil. The aerodynamics of rotating wing aircraft is even more complex because of the periodic change of the wing position with respect to the incoming flow. In this case, a sinusoidal variation of the angle of attack  in time (pitching airfoil) corresponds to an unsteady flow and the maximum lift condition is referred to as a \textit{dynamic stall}. In figure \ref{subfig.dynamic_stall} a pitching airfoil with a reduced frequency of 
$k=0.05$ and $\alpha\in\left[0^\circ,\ 20^\circ\right]$ is reported for the same VR12 airfoil.
The dynamic stall occurs at higher angles of attack compared to the steady counterpart, with  flow separation persisting when the angle of attack reduces (hysteresis effect).

The comparison of the numerical simulations are quite satisfactory in the proposed picture. The small shift of numerical and experimental results is very likely due to a not perfect angle of attack correction of wind tunnel data. Present limits of RANS and URANS methods however appear in the post-stall and deep-stall conditions dominated by a very massive separated flow. 

The static and dynamic stall characteristics have a critical role in defining performance of aircrafts, helicopters and wind turbines. Experimental studies are typically limited by the ability to achieve flight conditions in laboratory settings; numerical simulations, on the other hand, require the representation of the thin turbulent boundary layer developing on the airfoil surfaces, and several complex physical processes such as unsteadiness, laminar-turbulent transition and flow separation. The corresponding predictions are computationally expensive and extensive research continues to be devoted to the improvement and the assessment of the corresponding simulations. A summary of the state of the art for static stall is the AIAA high-lift prediction workshop \cite{Rumsey2019}.
 

As mentioned earlier, in this work we concentrate  on  static stall predictions and approach the problem using a machine learning technique. 

\begin{figure}[H]
    \centering
    \includegraphics[scale=0.025]{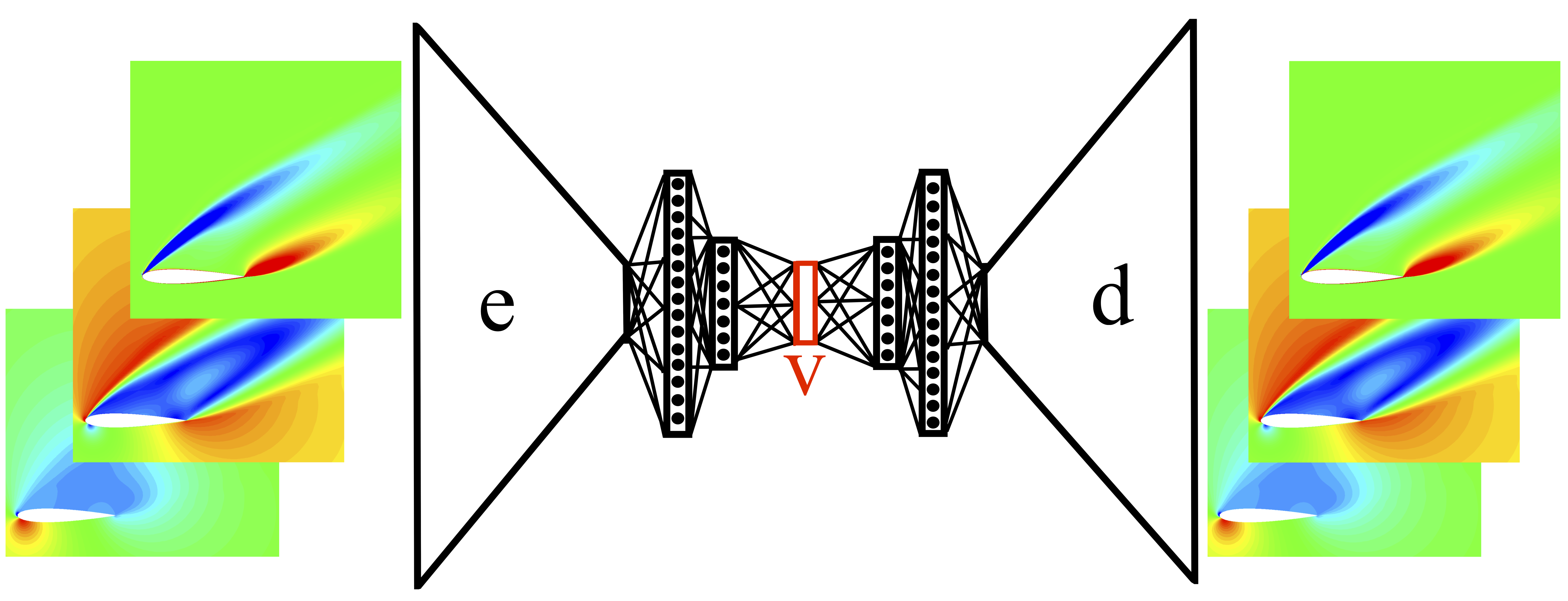}
    \caption{Simplified scheme of a convolutional autoencoder.}
    \label{fig.ae_scheme}
\end{figure}

Autoencoders (AE) are unsupervised deep-learning algorithm, which fundamentally target data compression, i.e. reducing the dimensionality of the input data  \cite{kookjin2020}.
The architecture of an autoencoder consists of two main elements, the encoder and the decoder. Giving an input dataset, AEs construct ({\it learn}) an equivalent  low dimensional representation, termed the \textit{latent space} using the \textit{encoder} portion of the network. On the other hand, The \textit{decoder}  rebuilds the input from the latent space. A simplified scheme of an autoencoder is reported in Fig. \ref{fig.ae_scheme} and summarized mathematically as:
\begin{equation}
    \vv = e\left(\vx\right); \qquad 
    \hat{\vx} = d\left(\vv\right); \qquad 
    \hat{\vx} = f\left(\vx\right) =  d\left(e\left(\vx\right)\right)  \ .
    \label{eq:code}
\end{equation}
where $e\left(\right)$ is the \textit{encoder} function, $d\left(\right)$ is the \textit{decoder} function, $f\left(\right)$ is the entire transformation \textit{encoder}+\textit{decoder}, $\vv$ is the vector of latent variables (also referred to as \textit{latent space}) and $\vx$ and $\hat{\vx}$ are respectively the input and its reconstruction.
\begin{equation*}
    \vx,\ \hat{\vx} \in \mathbb{R}^{n\times m\times c}; \qquad 
    \vv \in \mathbb{R}^p  \ .
\end{equation*}
where $n\times m\times c$ is the dimension of the input dataset and $p$ is the dimension of the \textit{latent space}.

In a Convolutional Autoencoder at least one of the layer of the network performs a convolution operation (denoted with $*$), which, in discrete form, is given by:
\begin{equation}
    s\left(t\right) = \left(x * w\right)\left(t\right) = \sum^\infty_{a=-\infty} x\left(a\right)w\left(t-a\right)
\end{equation}
where $x$ is the input, $w$ is the kernel and $s$ is the feature map.

Convolutional autoencoders have been already used for aerodynamic predictions, for instance
Bhatnagar et al. in \cite{Bhatnagar2019} showed their capability to reproduce the flow around airfoils with different shapes, angle of attack and Reynolds number. In their work they used a signed distance function (SDF) as input to the encoder and provided the velocity and pressure fields as output of the decoder. In this way it is possible to obtain the flow-field around a new geometry in real-time by using only geometrical information (the SDF).
Same methodology was adopted by Tangsali et al. in \cite{Tangsali2020}, but using a very rich training database (11,000+ CFD simulations), in order to demonstrate the generalizability of this approach for a wide range of airfoil geometries, Reynolds numbers and Mach numbers.

Rather than focusing on overall accuracy, Agostini \cite{Agostini2020}  focused on the interpretation of the latent space in the case of a flow around a cylinder. In his work, Agostini, showed the possibility to derive a low-dimensional dynamical model (in the latent space) by applying a clustering algorithm.


The performance of the autoencoder in predicting the fluid flow  around airfoils, shown in \cite{Tangsali2020} appears quite promising. However, the present approach is quite different: instead of constructing the autoencoder to generate the flow given a geometrical representation of the airfoil, we trained the autoencoder to reproduce the flow-field itself (as sketched in Fig. \ref{fig.ae_scheme}, input and output are the same) and then focused on the latent space representation to  generate new airfoils and flow-fields by using only the decoder.

The article is organized as follows:
first we provide a description of the training dataset construction;  then we analyze the latent variables of the autoencoder trained in the linear regime (airfoil operating in the linear part of the lift curve) for both images and raw data in input.
We performed a sensitivity analysis on the autoencoder architecture, hyperparameters and regularization method.
We present an investigation of the latent space for different values of its dimension to assess interpretability.
Finally, we investigate how to use the latent space with only the decoder in order to generate new synthetic airfoils and flow-fields by interpolation and extrapolation beyond the trained latent space.

\section{Database definition}
A critical element of data driven models is the dataset used for training. In the present work  we  consider numerical solutions of the Reynolds-Averaged Navier-Stokes (RANS) equations we obtained using the open-source software SU2.  Computations of the steady-state aerodynamic characteristics of
 NACA 4 digit airfoils are carried out. The free-stream conditions of Mach number and Reynolds number are fixed for all the  simulations: $M_\infty=0.15$, $Re_\infty=5\times10^6$.
 The  Spalart-Allmaras turbulence model is adopted and the flow is assumed \emph{fully turbulent}.

In particular, we constructed 2 databases:\begin{itemize}
    \item Linear case (33 RANS solutions), changing 2 parameters:\begin{itemize}
        \item[-] airfoil curvature (NACA 2412, 4412, 6412),
        \item[-] angle of attack $\alpha\in\left[-5^\circ,\ 5^\circ\right]$.
    \end{itemize}
    
    \item Non-linear case (124 RANS solutions), changing 2 parameters:\begin{itemize}
        \item[-] airfoil maximum thickness $t$ (NACA 0012, 0014, 0016 and 0018),
        \item[-] angle of attack $\alpha\in\left[0^\circ,\ 30^\circ\right]$.
    \end{itemize}
\end{itemize}

The first  dataset is used  to verify the autoencoder characteristics as it {\it learns}  linear aerodynamics (low-angle of attack) and it is described in section \ref{sec.:sens_analysis}.
The second dataset is used for the investigation of the airfoil static stall discussed in section \ref{sec.LatSp_invest}.

The datasets include a subset of the field variables computed as part of the RANS solutions; specifically we included the 
pressure coefficient $C_p$, the vorticity $\omega$ and the Mach number $M$ as they are directly relevant to the stall detection.  $C_p$ is the critical contribution to the aerodynamic forces and therefore critical to establish the flow regime and the loss of lift;
 $\omega$ provides information about the boundary layers and the presence and  position of an eventual separation points on the surface of the airfoil; $M$ is correlated to the velocity field and yields indication of the flow separation and the wake structure. In addition to field quantities, we also include geometrical information in the dataset to distinguish different airfoils; specifically we included the $x$ and $y$ coordinates of the mesh grid points.

\begin{figure}[H]
    \centering
    \includegraphics[scale=0.3]{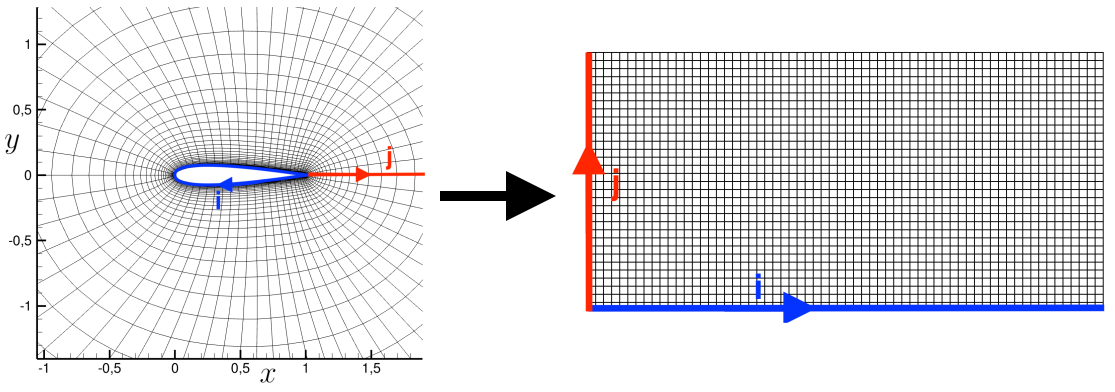}
    \caption{$ij$ mapping.}
    \label{img.ijmapping}
\end{figure}

The SU2 computations were carried out using a structured O-mesh (as reported in Fig.\ref{img.ijmapping}); the field variables are organized in matrices using the  $i$ and $j$ indexes of the mesh. The $i$ index is a curvilinear coordinate on the airfoil surface and it corresponds to the horizontal axis of the encoded mapping, while the $j$ index tracks the radial direction away from the airfoil and corresponds to the vertical axis encoded mapping. This input representation has also the advantage of amplifying the boundary layer as $j$ does not follow the mesh clustering at the airfoil surface  (see Fig.\ref{img.ijmapping})); this in turns enables us to highlight the boundary layers in the training step. 
In summary, the input to the convolutional autoencoder consists of 2D matrices ($ij$-mappings) each with 5 channels ($c_p$, $\omega$, $M$, $x$ and $y$), and the variables are scaled between 0 and 1. The datasets are composed of either images (RGB channels) or raw solutions (matrix of
floating points). Therefore, the number of input channels is 15 for the images (3 RGB $\times$ 5 variables) and 5 for the raw data (1 channel $\times$ 5 variables).
\section{Autoencoder sensitivity analysis}\label{sec.:sens_analysis}

\subsection{Architecture}
As a preliminary step of the present investigation we carried out an extensive  analysis to assess  the effect of the network architecture (number of layers, activation functions, etc.) and the training strategy and related  hyperparameters. As a trade-off between ease of training and accuracy, we focused on the following structure:

\begin{itemize}
        \item \textbf{Encoder}:\begin{itemize}
            \item[-] 4 convolutional layers.
            \item[-] 2 linear layers.
            \item[-] ReLU activation functions.
        \end{itemize}
        \item \textbf{Latent space dimension}: { $p=1,2,3 \ .$}
        \item \textbf{Decoder}:\begin{itemize}
            \item[-] 2 linear layers.
            \item[-] 4 transposed convolutional layers.
            \item[-] ReLU activation functions; Sigmoid for the last layer.
        \end{itemize}
\end{itemize}

We trained the autoencoder using a completely randomized dataset with a batch size equal to 2, the Adam optimizer and the Mean Squared Error (MSE) as loss function.
The algorithm is implemented using the \textit{Pytorch} library \cite{Pytorch2019} and trained on a single GeForce GTX Titan X GPU.

The latent space dimension $p$ is the most important hyperparameter of an autoencoder.
By increasing $p$ it is possible to reach a certain level of tolerance in the training process with less epochs. However,  a high dimensional latent space can lead to overfitting and ultimately,  poor performance on new unseen cases.
For our purposes, another critical design goal  is  interpretability: a very high dimensional latent space limits our ability to  extract  physical understanding from the latent variables. In the results sections We focus on extremely low values of ($p \le 3$) and in many cases we have achieved error tolerances that are comparable to   richer latent representations. 

\begin{figure}[H]
    \centering
    \includegraphics[scale=0.3]{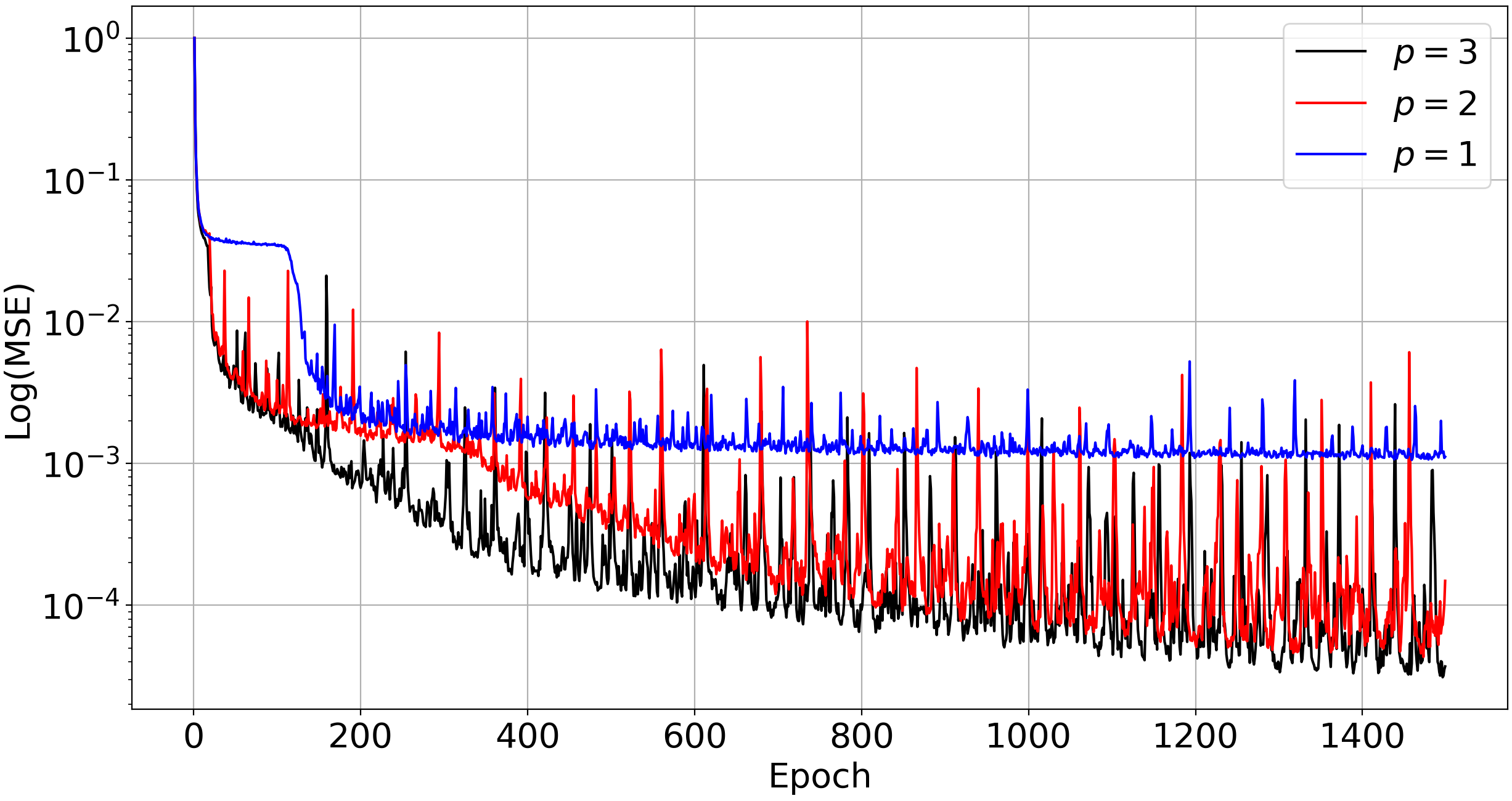}
    \caption{Comparison of the convergence training procedures for $p=1,\ 2$ and $3$ using the non-linear dataset.}
    \label{fig:training_convergence}
\end{figure}

As an example, in Fig. \ref{fig:training_convergence} we report the convergence of the training procedure for the same autoencoder and dataset (the non-linear regime) for $p=1,\ 2$ and $3$. The results illustrate that $p=2$ and $p=3$ reach a reasonably close level of accuracy in the training, while $p=1$ is clearly insufficient in this case.

\begin{figure}[H]
    \centering
    \includegraphics[scale=0.3]{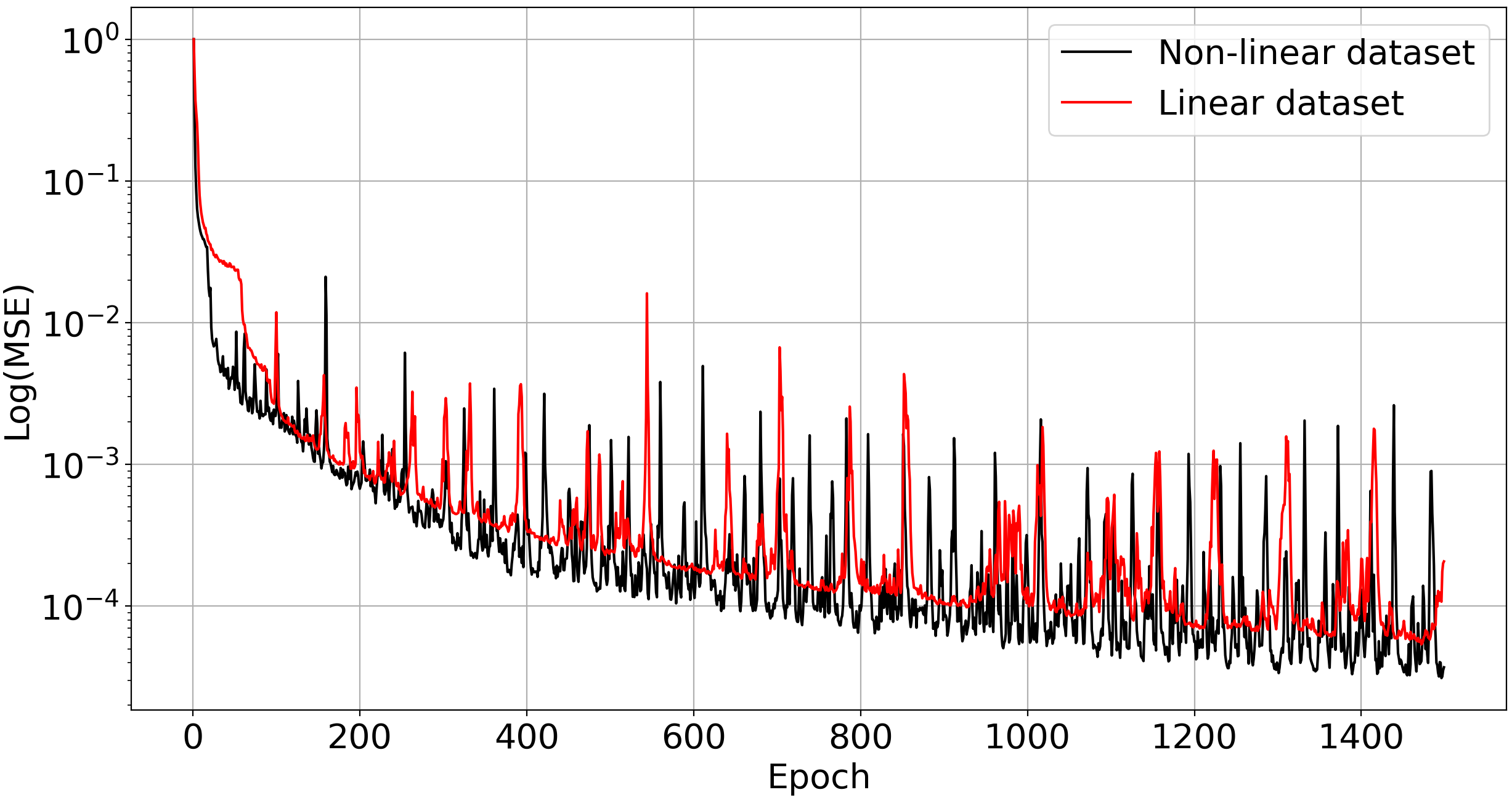}
    \caption{Comparison of the convergence training procedures for the non-linear and linear datasets ($p=3$).}
    \label{fig:training_convergence_lin}
\end{figure}

In Fig. \ref{fig:training_convergence_lin} we compare the convergence of the training procedure for the two datasets (linear and non-linear) using the same autoencoder with $p=3$. The two cases show the same rate of decay of the loss function. The comparison is done using the same training parameters (e.g. learning rate) for both linear and non-linear case without any tuning to speed-up the training process.

In our specific case, 2 parameters are varying in the database: the airfoil thickness and the angle of attack. For this reason, at least 2 latent variables are necessary and sufficient, to describe our dataset. For our analysis we decided to use 3 latent variables that is a good compromise between latent space interpretability and training speed performance.

\subsection{Input type: images or raw data?}
As a  first step we investigate  the effect of the  type of input  for the convolutional autoencoder.  Images are widely used in this context and therefore, we start by investigating  if images provide a sufficient quantitative representation of the physical solutions.
We performed a sensitivity analysis by training the autoencoder with the linear dataset and investigated the behavior of the latent variables (with $p=3$).

\begin{figure}[H]
\setcounter{subfigure}{0}
    \centering
    \subfloat[Image, contour levels: 50]{\includegraphics[scale=0.3]{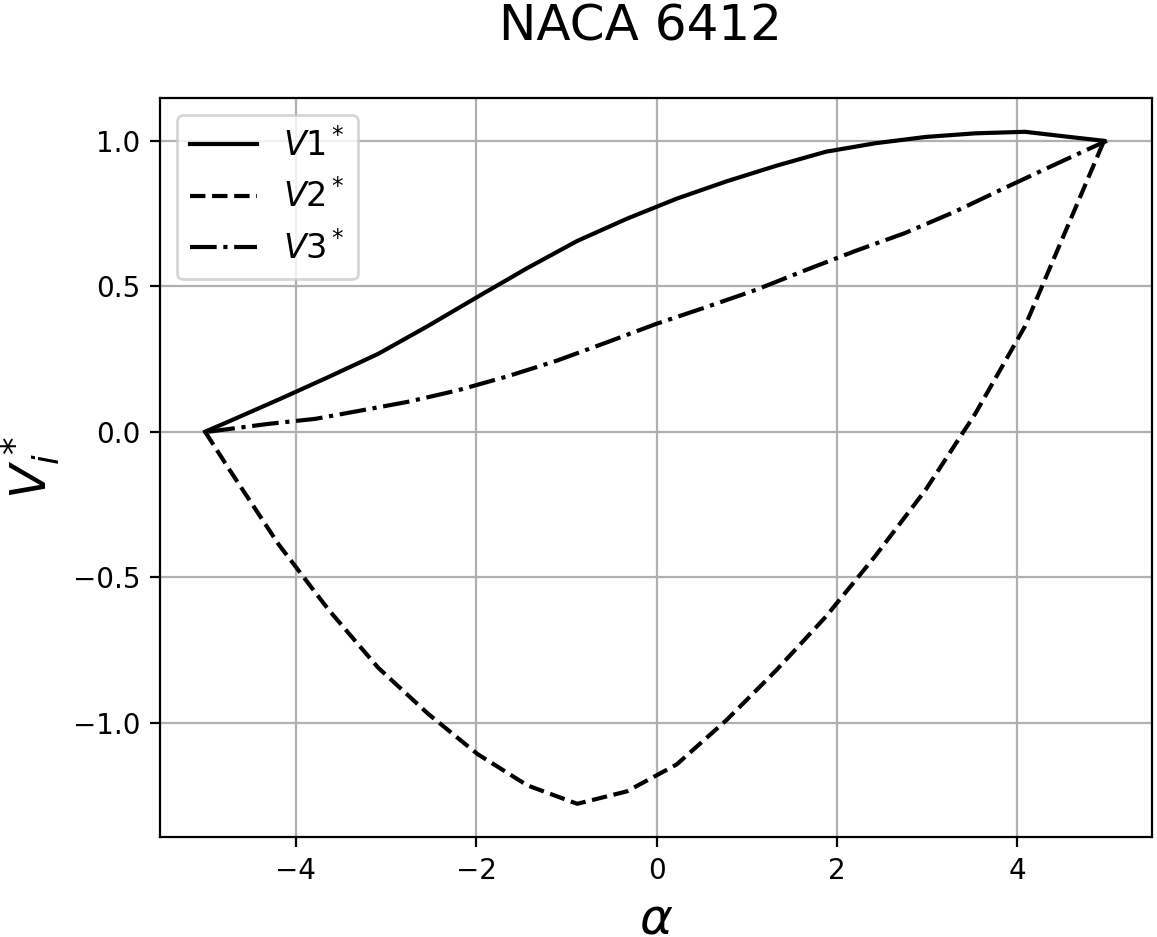}\label{subFig.lvl50}}
    \subfloat[Image, contour levels: 500]{\includegraphics[scale=0.3]{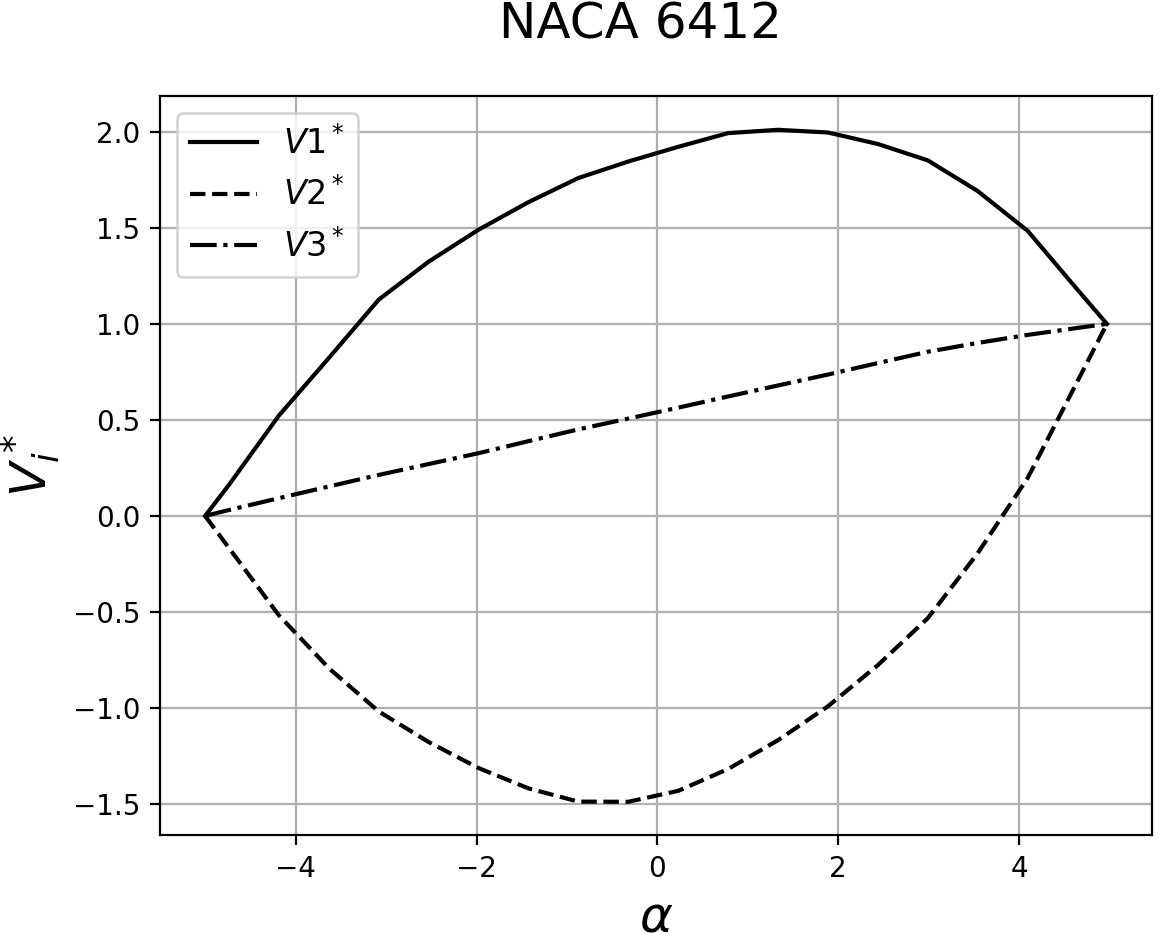}\label{subFig.lvl500}}
    \subfloat[Image, contour levels: 5000]{\includegraphics[scale=0.3]{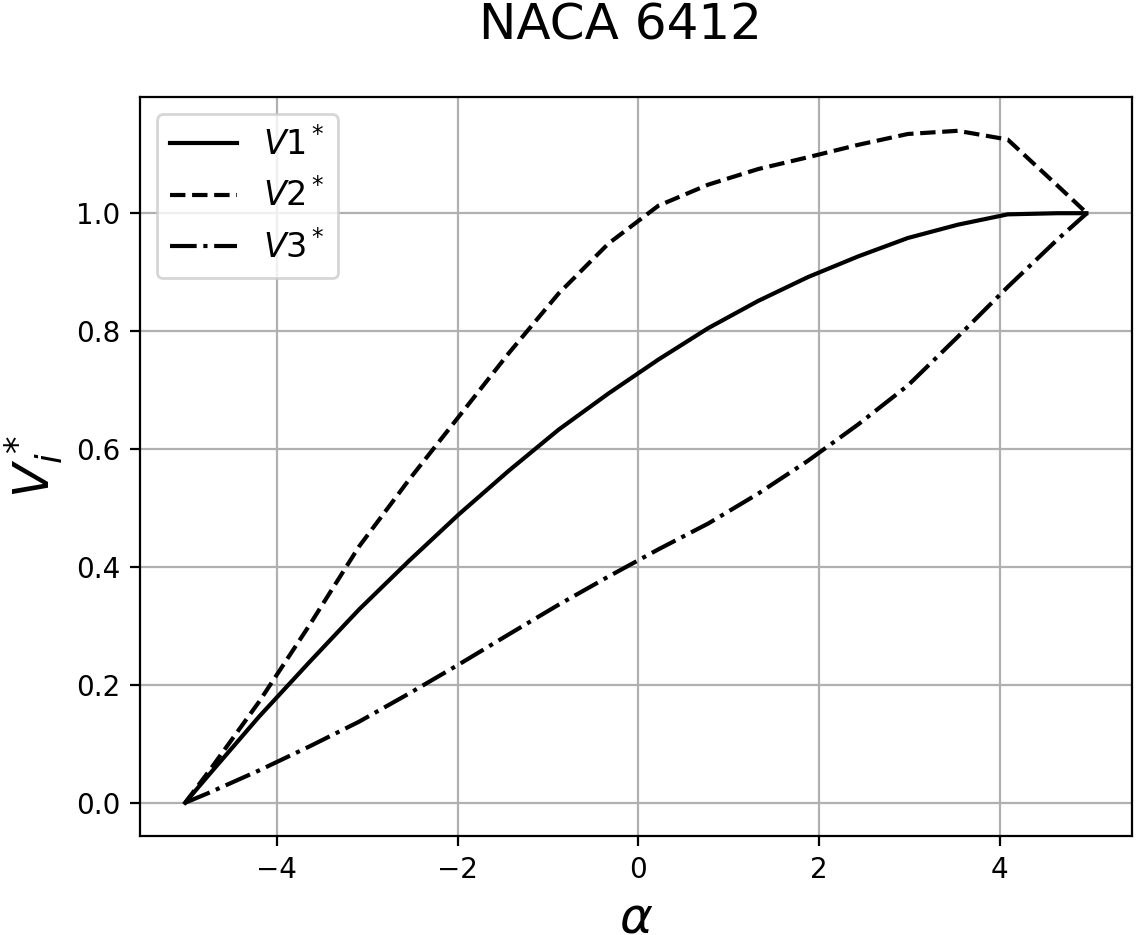}\label{subFig.lvl5000}}
    \subfloat[Raw data]{\includegraphics[scale=0.3]{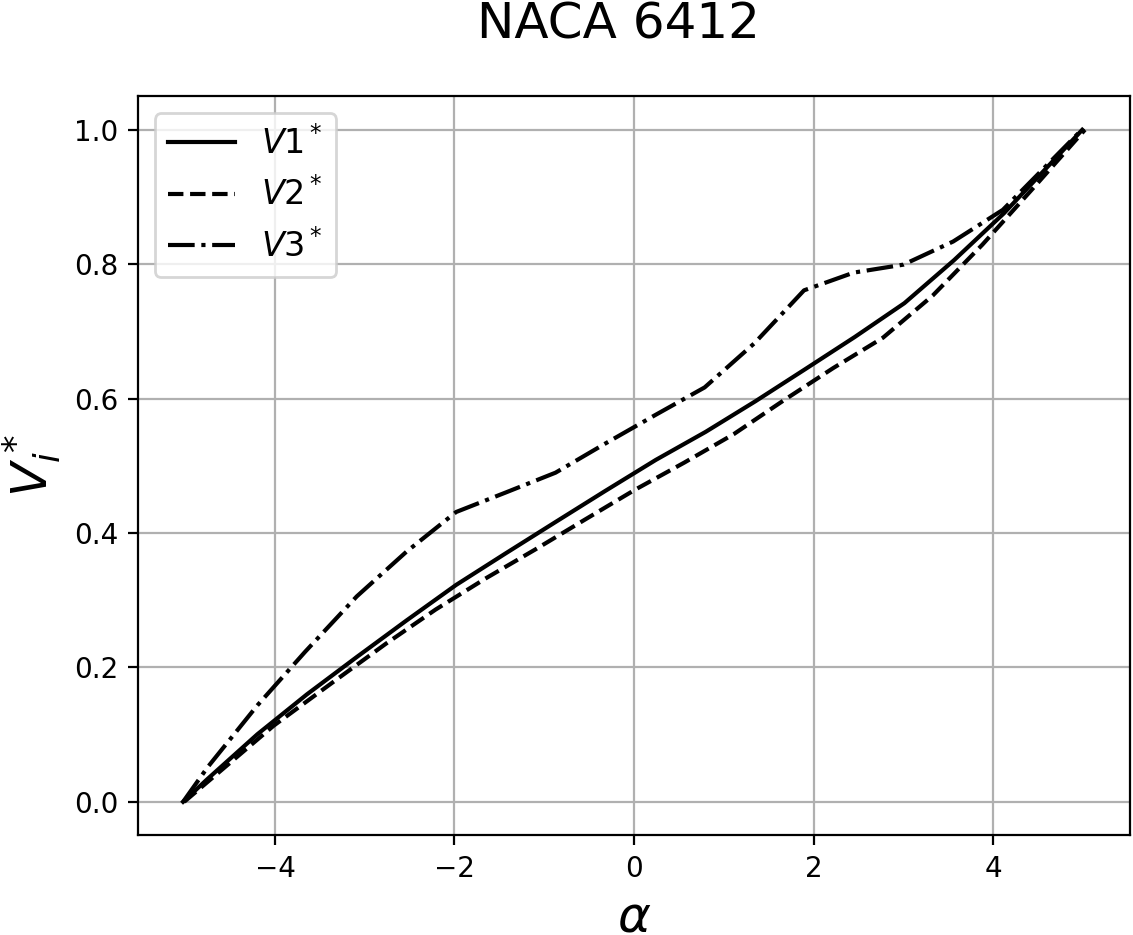}\label{subFig.raw}}
\caption{Latent variables extracted by the AE (normalized between 0 and 1). Linear case: NACA 6412.}
\label{fig.latVar_lin}
\end{figure}

Fig. \ref{subFig.lvl50}, \ref{subFig.lvl500} and \ref{subFig.lvl5000} show normalized latent variables learned by the autoencoder trained with images constructed using  different contour levels: 50, 500 and 5000 respectively. These are also compared to using the raw data directly (Fig. \ref{subFig.raw}). It is clear that 
the latent variables reported as a function of the angle of attack are quite sensitive to the input format. It is expected that in these conditions the latent space represents closely the variability expressed by the angle-of-attack. The results confirm that increasing the number of contour levels in the image (e.g. increasing the precision in representing the data) leads to a better correlation between the latent variables and the expected variability.  Using the raw data as input leads to a collapse of the latent space and effectively a purely linear response to the changes in the angle-of-attack. 



\begin{figure}[H]
\setcounter{subfigure}{0}
    \centering
    \subfloat[Image, contour levels: 50]{\includegraphics[scale=0.30]{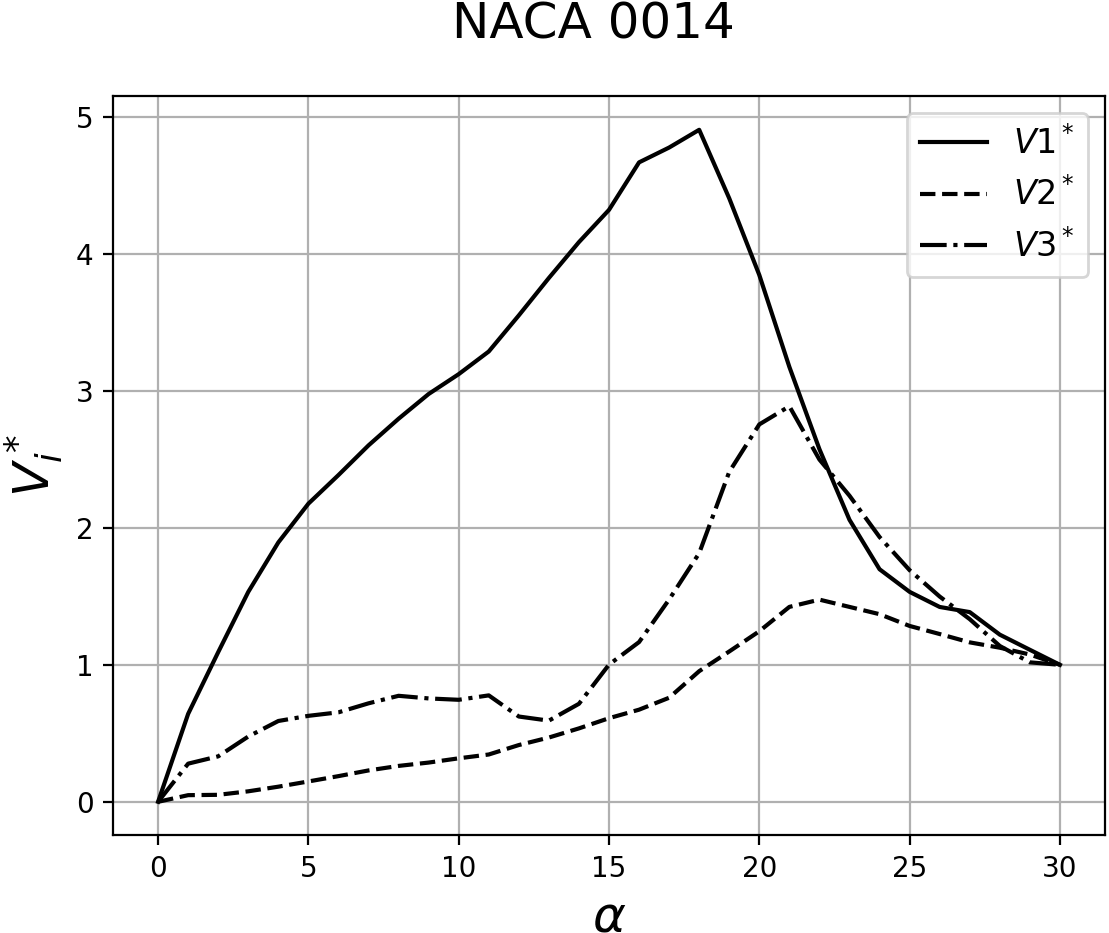}}
    \subfloat[Image, contour levels: 500]{\includegraphics[scale=0.30]{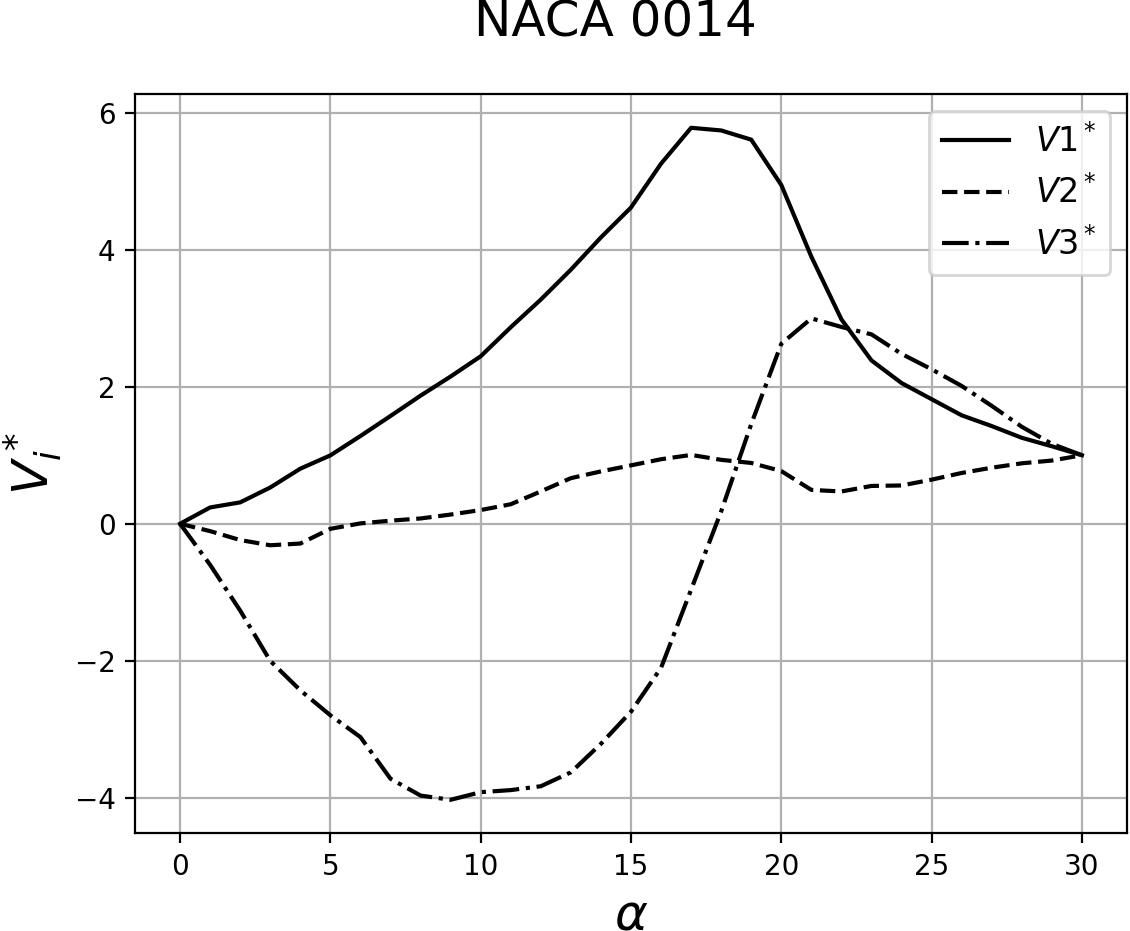}}
    \subfloat[Image, contour levels: 5000]{\includegraphics[scale=0.30]{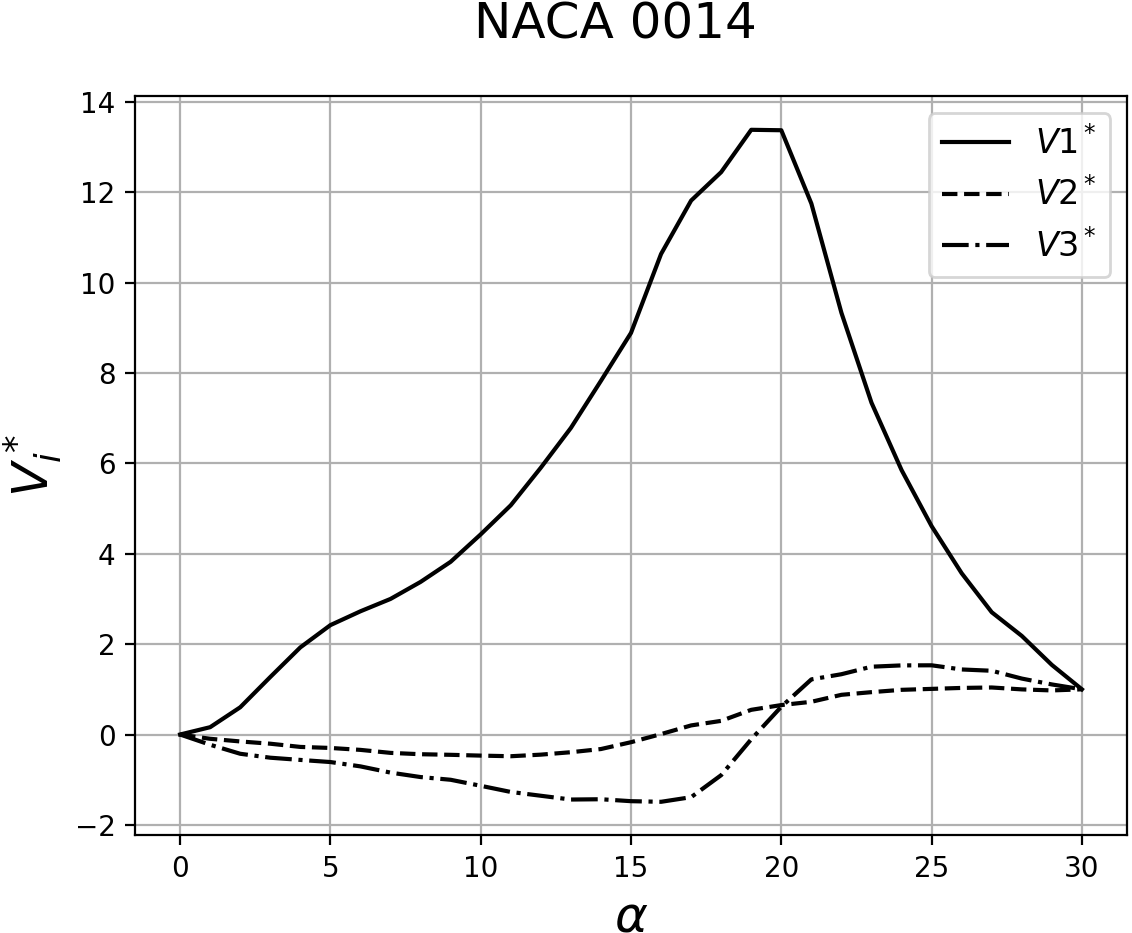}}
    \subfloat[Raw data]{\includegraphics[scale=0.30]{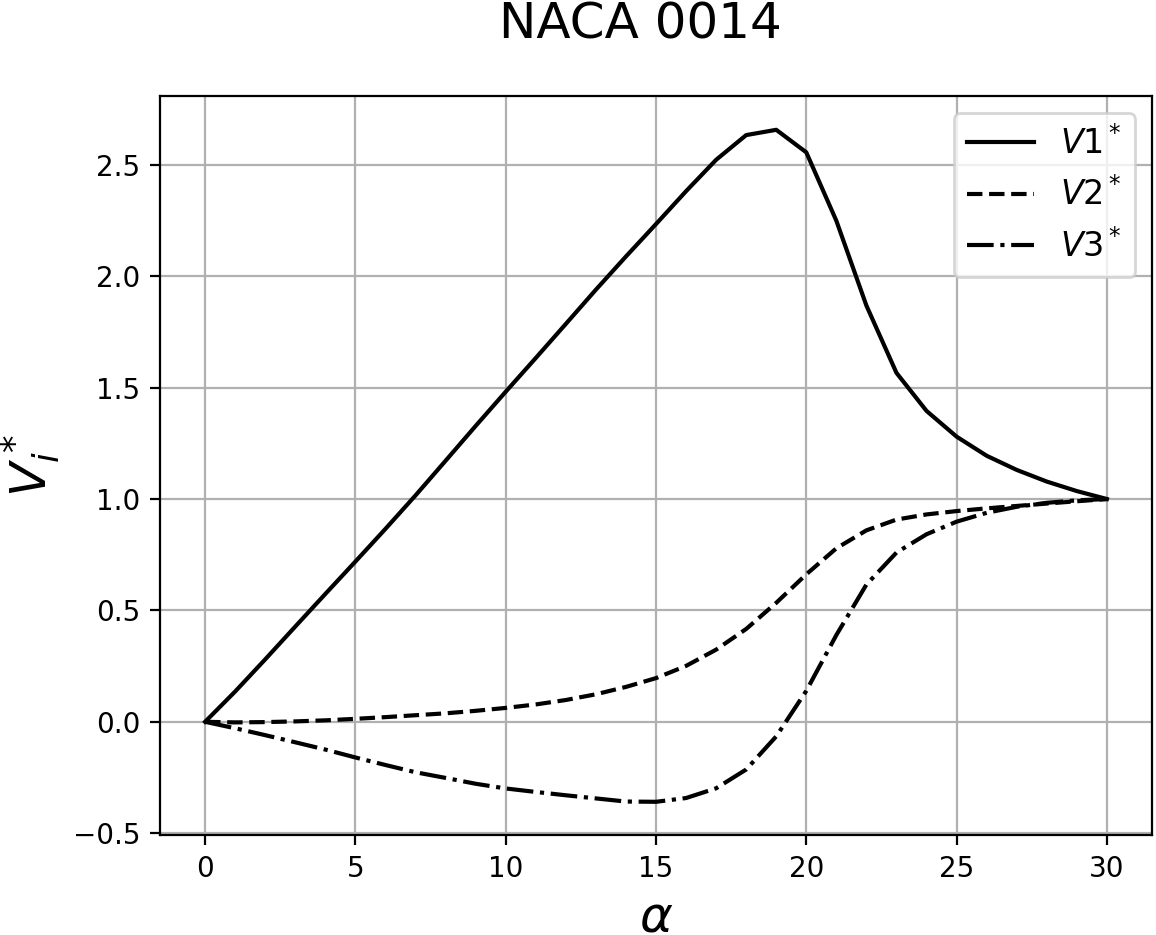}}
\caption{Latent variables provided by the AE (scaled between 0 and 1). Non-linear case: NACA 0014.}
\label{fig.latVar_nonLin}
\end{figure}

Fig. \ref{fig.latVar_nonLin} shows the same sensitivity analysis carried out in the non-linear regime.
As before, we  notice a sensitivity of the latent space to the number of contour levels. Moreover, the raw data in input produced latent variables which are linear for low angle-of-attack consistently with what shown previously.
All the results reported in what follows are obtained by using raw data in input.

\subsection{Regularization}
\begin{figure}[H]
    \setcounter{subfigure}{0}
    \centering
    \subfloat[No regularization]{\includegraphics[scale=0.45]{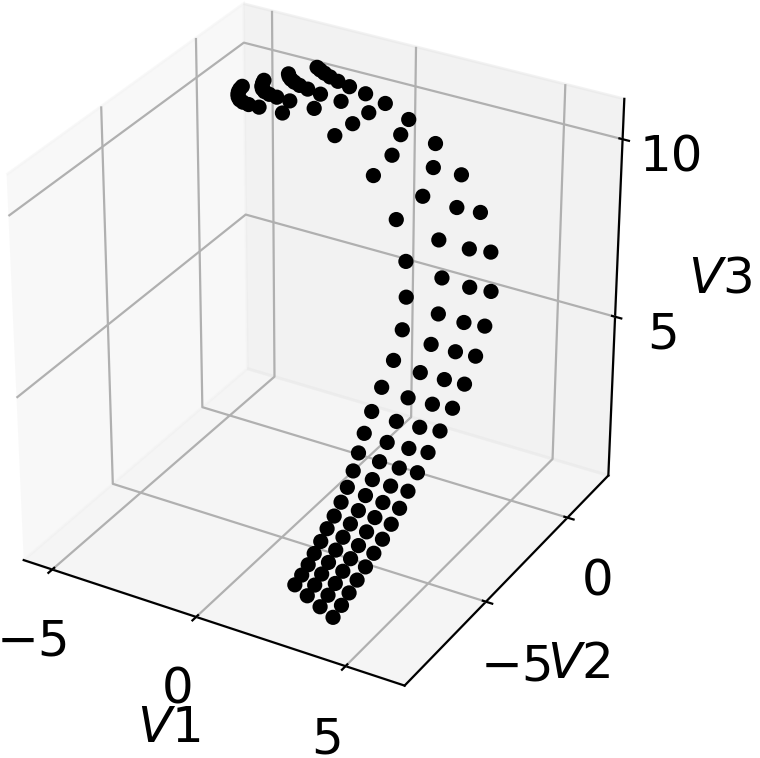}\label{subfig.noReg}}
    \subfloat[Latent space regularization]{\includegraphics[scale=0.45]{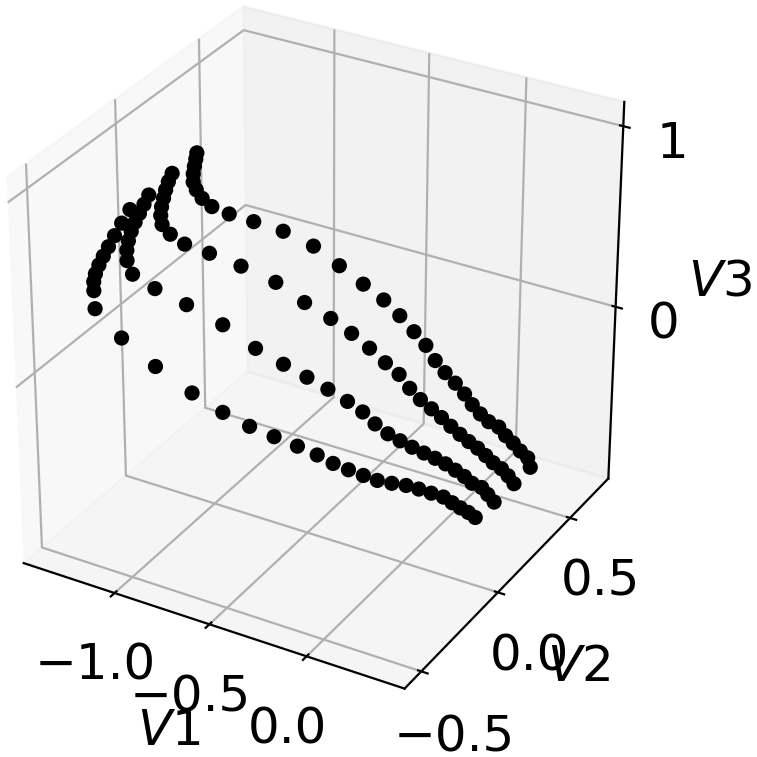}\label{subfig.LatSpReg}}
    \subfloat[Encoder weights regularization]{\includegraphics[scale=0.45]{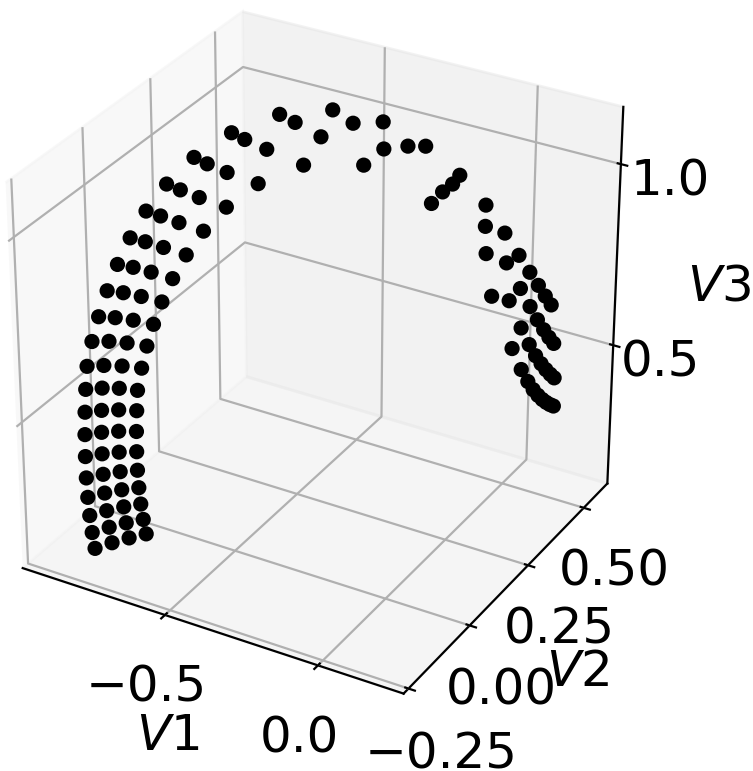}\label{subfig.EncReg}}

\caption{Comparison of latent spaces for $p=3$ changing the regularization technique. Autoencoder trained with the non-linear data-set.}
\label{fig.RegAn}
\end{figure}

As described by Newson et al. in \cite{Newson2020}, different regularization methods can lead to different behavior of the latent variables and different reconstruction performance by the autoencoder. 
We studies the sensitivity of the present algorithm to using $L_2$ norm regularization on the encoder and/or decoder weights and on the latent variables. Eventually we found that the $L_2$ norm regularization applied only to the encoder weights leads to a tidy and compact latent space which is amenable to interpretation. As a comparison of the various types of regularization we report the latent variables obtained using the non-linear regime dataset and $p=3$ (Fig. \ref{fig.RegAn}): the results are not vastly different, although it is clear that regularizing the   the encoder weights leads to a  compact latent space.



The latent space regularization in Fig. \ref{subfig.LatSpReg} is obtained by applying a penalization on the $L_2$ norm of the latent variables, while the encoder weights regularization in Fig. \ref{subfig.EncReg} is an $L_2$ norm of the weights of the Encoder.

\section{Latent space investigation}\label{sec.LatSp_invest}
Based on the sensitivity analysis illustrated earlier, we trained
the autoencoder
applying the $L_2$ norm regularization on the encoder weights.
The non-linear regime dataset is considered for the investigation of the latent space, considering raw data and variables normalized between 0 and 1. The input channels are $C_p$, $M$, $\omega$, $x$ and $y$.

\begin{figure}[H]
    \setcounter{subfigure}{0}
    \centering
    \subfloat[$p=1$ latent space]{\includegraphics[scale=0.35]{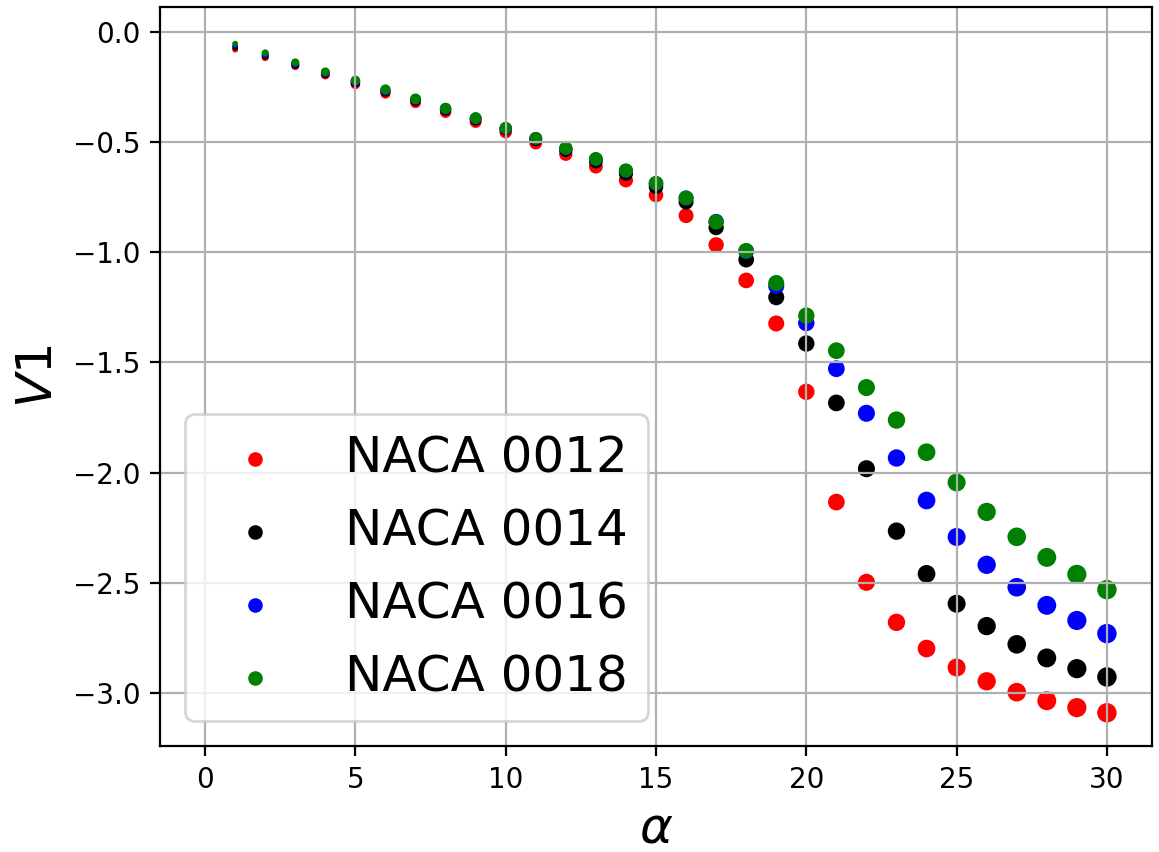}\label{subfig.p1}}
    \subfloat[$p=2$ latent space]{\includegraphics[scale=0.35]{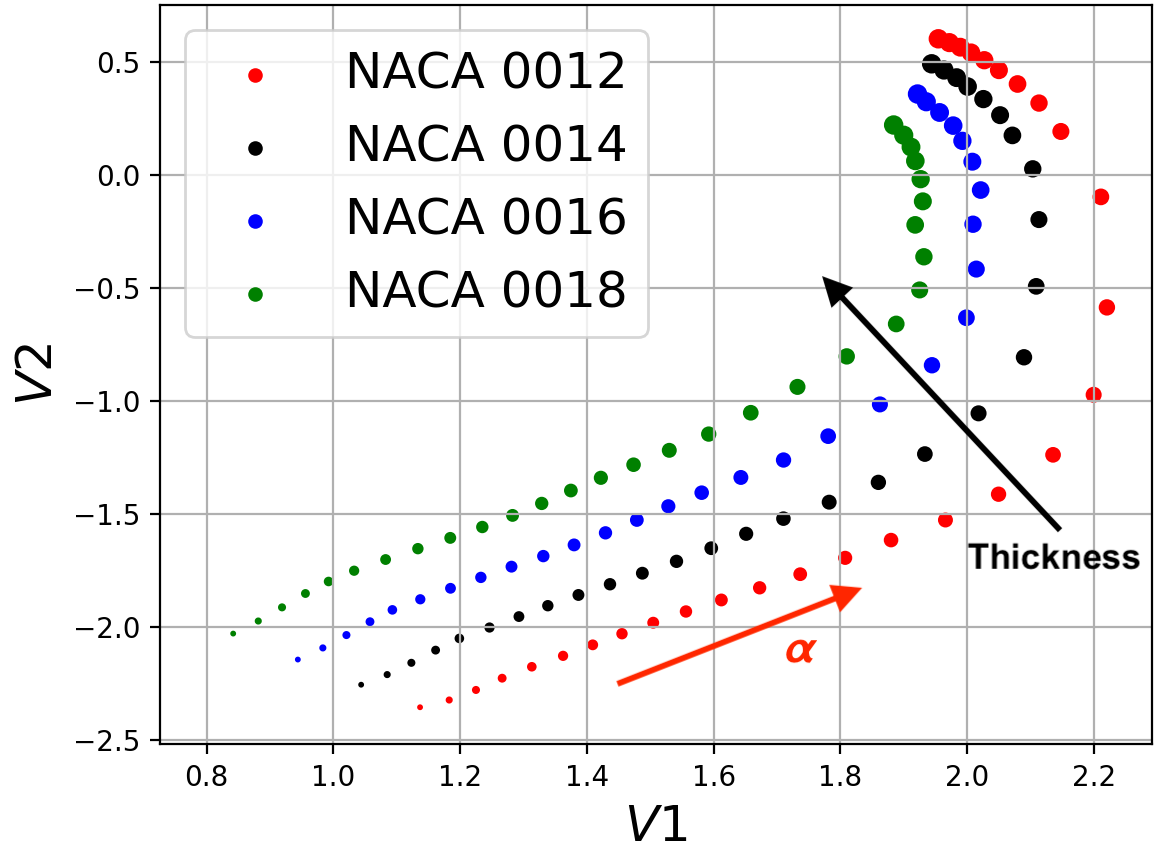}}
    \subfloat[ $p=3$ latent space]{\includegraphics[scale=0.45]{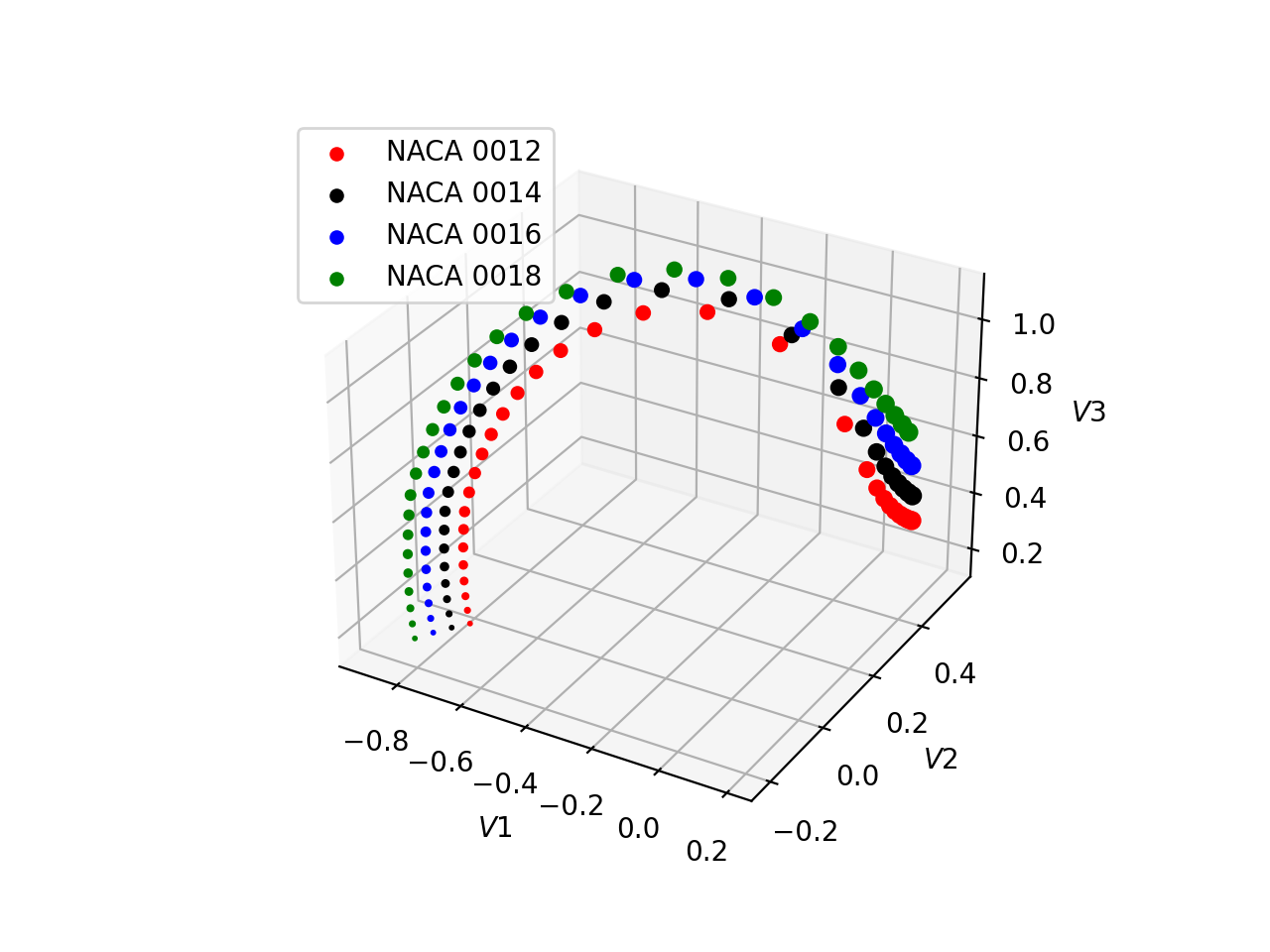}}

\caption{$p=1$, $2$, $3$ latent spaces learned by the convolutional autoencoder trained with the non-linear dataset.}
\label{fig.latSpaces}
\end{figure}

Fig. \ref{fig.latSpaces} shows the latent variables for $p=1$, $p=2$ and $p=3$ learned by the autoencoder. The dataset consists of RANS computations of the flow around symmetric airfoil of different thickness at various angles of attack; effectively two free input parameters (thickness $t$ and $\alpha$) span the entire database. 

By mapping the latent space to the free input parameters is clear that the autoencoder is correctly ordering the solutions although $t$ and $\alpha$ are not explicitly given as inputs and the training process is  randomized. 


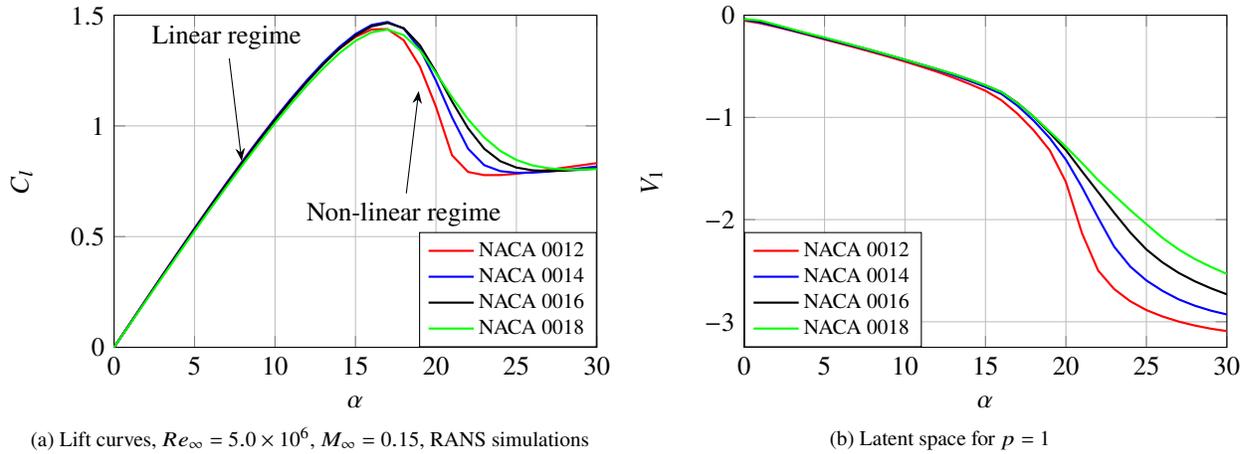
\begin{figure}[H]
    \centering
\subfloat[Lift curves, $Re_\infty=5.0\times10^6$, $M_\infty=0.15$, RANS simulations]{
\begin{tikzpicture}
    \pgfplotsset{every axis legend/.append style={nodes={scale=0.8}, at={(1,0)},anchor=south east}}
    
    \begin{axis}[height=6cm, width=8cm, grid, xmin=0, xmax = 30,ymin=0,ymax=1.5,xlabel=$\alpha$, ylabel=$C_l$]
    \node at (180,60) (nodeA) {Non-linear regime};
    \node at (190,120) (nodeB) {};
    
    \draw [-{Stealth}] (nodeA) -- (nodeB);
    
    \node at (70,140) (nodeA) {Linear regime};
    \node at (80,80) (nodeB) {};
    
    \draw [-{Stealth}] (nodeA) -- (nodeB);

    \addplot[thick,solid, color=red, mark size = 3pt]
    table [x expr=\thisrowno{0}, y expr=\thisrowno{1}] {Plots/AeroForces_n0012.dat};\addlegendentry{NACA 0012}
    \addplot[thick,solid, color=blue, mark size = 3pt]
    table [x expr=\thisrowno{0}, y expr=\thisrowno{1}] {Plots/AeroForces_n0014.dat};\addlegendentry{NACA 0014}
    \addplot[thick,solid, color=black, mark size = 3pt]
    table [x expr=\thisrowno{0}, y expr=\thisrowno{1}] {Plots/AeroForces_n0016.dat};\addlegendentry{NACA 0016}
    \addplot[thick,solid, color=green, mark size = 3pt]
    table [x expr=\thisrowno{0}, y expr=\thisrowno{1}] {Plots/AeroForces_n0018.dat};\addlegendentry{NACA 0018}
    \end{axis}
\end{tikzpicture}
\label{fig.liftCurve}
}
\subfloat[Latent space for $p=1$]{
\begin{tikzpicture}
    \pgfplotsset{every axis legend/.append style={nodes={scale=0.8}, at={(0,0)},anchor=south west}}
    
    \begin{axis}[height=6cm, width=8cm, grid, xmin=0, xmax = 30,ymin=-3.25,ymax=0,xlabel=$\alpha$, ylabel=$V_1$]

    \addplot[thick,solid, color=red, mark size = 3pt]
    table [x expr=\thisrowno{0}, y expr=\thisrowno{1}] {Plots/LatVar_1D.dat};\addlegendentry{NACA 0012}
    \addplot[thick,solid, color=blue, mark size = 3pt]
    table [x expr=\thisrowno{0}, y expr=\thisrowno{2}] {Plots/LatVar_1D.dat};\addlegendentry{NACA 0014}
    \addplot[thick,solid, color=black, mark size = 3pt]
    table [x expr=\thisrowno{0}, y expr=\thisrowno{3}] {Plots/LatVar_1D.dat};\addlegendentry{NACA 0016}
    \addplot[thick,solid, color=green, mark size = 3pt]
    table [x expr=\thisrowno{0}, y expr=\thisrowno{4}] {Plots/LatVar_1D.dat};\addlegendentry{NACA 0018}
    \end{axis}
\end{tikzpicture}
}
\caption{Comparison of the lift curves of the airfoils of the non-linear data-set: NACA 0012, 0014, 0016 and 0018 and the latent variable learned by the autoencoder with the same dataset for $p=1$.}
\end{figure}

Fig. \ref{subfig.p1}  shows that $p=1$ is not sufficient to describe the entire dataset;  for $\alpha<10^\circ$, one value of the latent variable corresponds to multiple airfoils. However, this case illustrates an important finding that connects the latent variables to the physics of the phenomenon. 
The autoencoder infers that symmetrical airfoils have the same behaviour at low angle of attacks, while in the non-linear regime the thickness plays an important, differentiating role. 
This is confirmed by the analysis of the lift curves extracted from the RANS solutions (Fig. \ref{fig.liftCurve}), which completely overlap for $\alpha < 10^\circ$. 

\begin{figure}[H]
    \centering
   \begin{tikzpicture}
    \pgfplotsset{every axis legend/.append style={nodes={scale=0.8}, at={(1,0)},anchor=south east}}
    
    \begin{axis}[height=6cm, width=8cm, grid, xmin=0, xmax = 3.25,ymin=0,ymax=1.625,xlabel=$-V_1$, ylabel=$C_l$]

    \addplot[only marks, color=red, mark size = 1pt]
    table [x expr=-\thisrowno{1}, y expr=\thisrowno{0}] {Plots/V1_vs_Cl.dat};\addlegendentry{NACA 0012}
    \addplot[only marks, color=blue, mark size = 1pt]
    table [x expr=-\thisrowno{3}, y expr=\thisrowno{2}] {Plots/V1_vs_Cl.dat};\addlegendentry{NACA 0014}
    \addplot[only marks, color=black, mark size = 1pt]
    table [x expr=-\thisrowno{5}, y expr=\thisrowno{4}] {Plots/V1_vs_Cl.dat};\addlegendentry{NACA 0016}
    \addplot[only marks, color=green, mark size = 1pt]
    table [x expr=-\thisrowno{7}, y expr=\thisrowno{6}] {Plots/V1_vs_Cl.dat};\addlegendentry{NACA 0018}
    \end{axis}
\end{tikzpicture}

    \caption{Lift coefficient $C_l$ as function of the latent variable learned by the autoencoder for $p=1$.}
    \label{fig:correlation}
\end{figure}
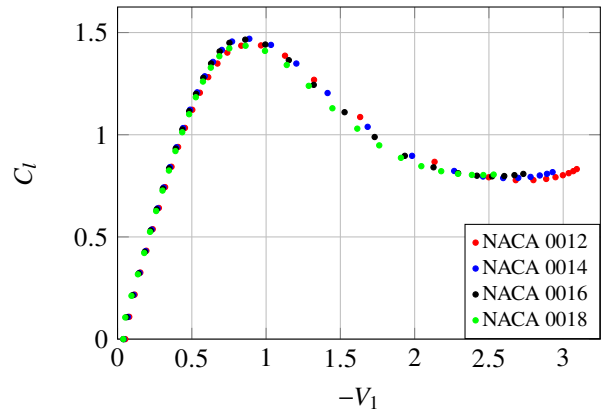

The existance of a strong correlation between the lift coefficient $C_l$ and the latent variable $V_1$ for $p=1$ is clearly visible also in Fig. \ref{fig:correlation}, where the $C_l$ is reported as function of $V_1$.
The different curves for the 4 airfoils collapse almost perfectly.
In particular, the correlation between $C_l$ and $V_1$ is greater than $90\%$ in the linear regime.

\section{Generating synthetic airfoils}
The decoder element of the present algorithm can be used to quickly generate new {\it synthetic} solutions by modifying the values of the latent variables. As an final test we investigated the ability of the autoencoder to generate new airfoils and their aerodynamic characteristics. 

\subsection{Latent variables translation}
\begin{figure}[H]
\setcounter{subfigure}{0}
    \centering
    \includegraphics[scale=0.4]{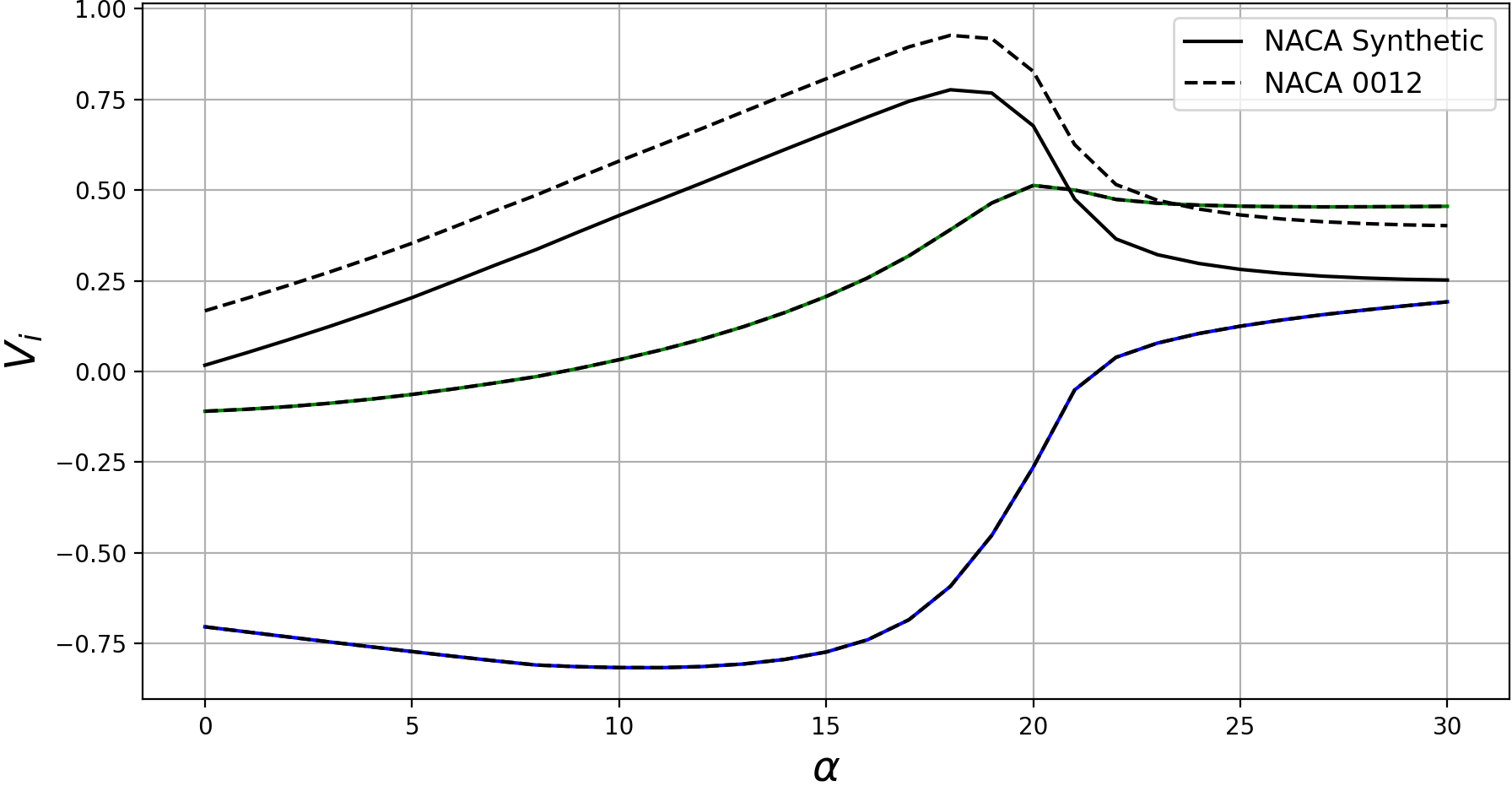}
    \caption{Translation of latent variables as function of the angle of attack $\alpha$. Comparison with the latent variables of the NACA 0012 airfoil. $V_1$: {\color{blue}---}; $V_2$: {\color{green}---}; $V_3$: --- .}
    \label{Fig.latVarTra_alpha}
\end{figure}

A simple modification to the latent variables is first attempted.
We considered the latent variables of the NACA 0012 as baseline. Fig. \ref{Fig.latVarTra_alpha} shows the variables as function of the angle of attack for the NACA 0012. We apply a translation of one of the component $V_3$ by $~10\%$.

\begin{figure}[H]
\setcounter{subfigure}{0}
    \centering
    \includegraphics[scale=0.07]{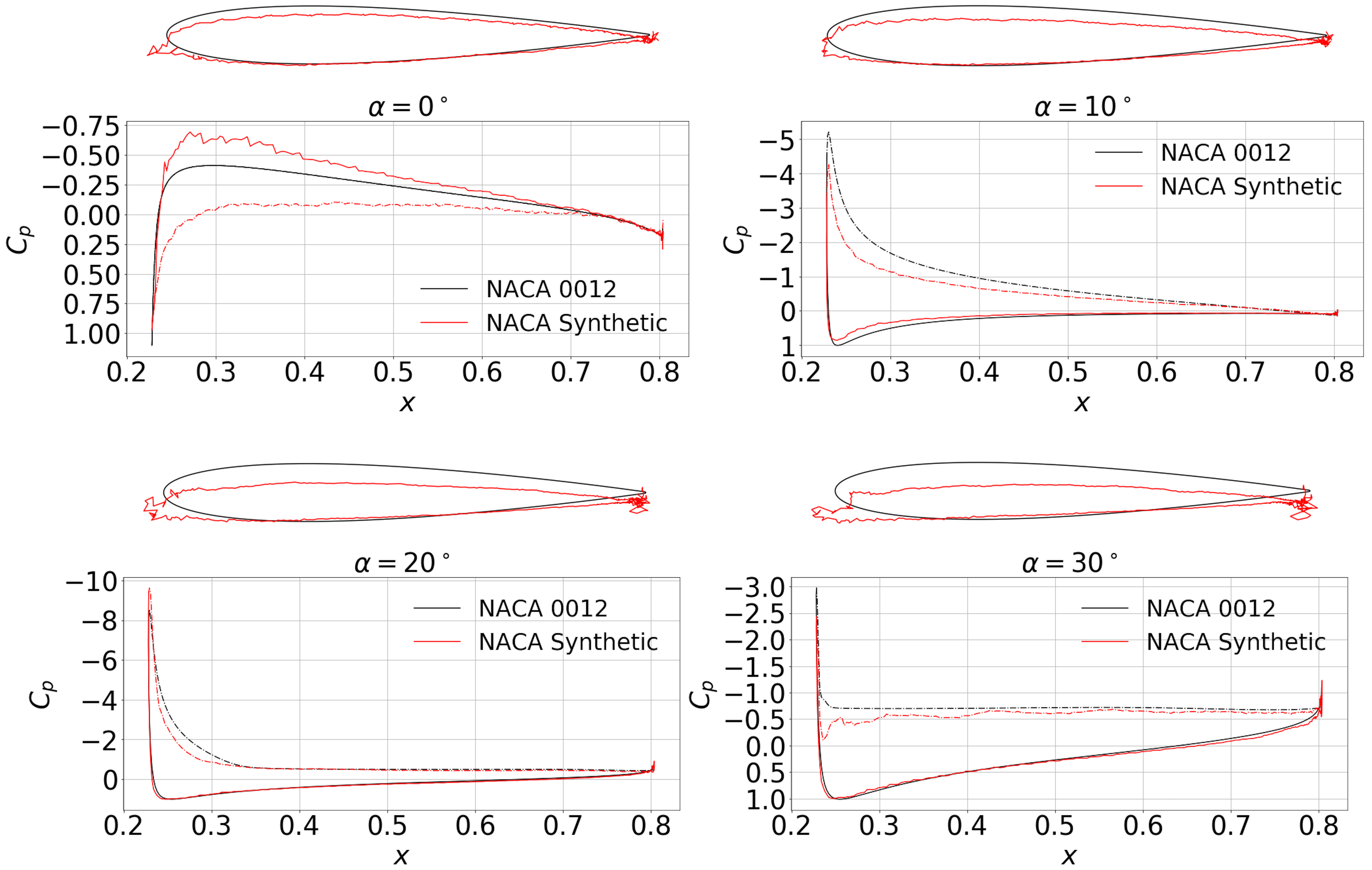}
    \caption{Comparison between the decoder output by giving in input the translated latent variables and the NACA 0012 baseline for $\alpha=0^\circ,\ 10^\circ,\ 20^\circ$ and $30^\circ$.}
    \label{Fig.latVarTra_cp}
\end{figure}

In Fig. \ref{Fig.latVarTra_cp}, the output of the decoder with the new latent variables is compared with the NACA 0012 baseline in terms of pressure distributions and airfoil geometry for 4 selected angles of attack ($\alpha=0^\circ,\ 10^\circ,\ 20^\circ$ and $30^\circ$). As a first observation, it is clear that the synthetic geometry changes with the angle of attack, which can be interpreted as an indication that changes in the latent space variables cannot be arbitrary.

In the $C_p$ plots (Fig. \ref{Fig.latVarTra_cp}) the continuous lines refer to the lower surface of the airfoil, while the dashed lines to the upper surface. The decoder extracts an airfoil and a $C_p$ distribution for $\alpha=0^\circ$  that corresponds to a negative angle of attack for a symmetric airfoil.


The $C_p$ curves and the airfoil geometries reported in Fig. \ref{Fig.latVarTra_cp} are obtained by extracting the first row of the reconstructed channels ($C_p$, $x$ and $y$). The AE is trained giving in input $170\times512\times5$ tensors, therefore the airfoil geometry is the reconstruction of 2 vectors ($x$ and $y$ coordinates) of $512$ elements each.
This means that the airfoil geometry is only the 0.23\% of the information that the AE is reproducing.
In addition, the poor quality of the leading and trailing edge reconstructions are due to the presence of a thickening of the mesh grid in these regions, so the values of the coordinates are very close to each other and a small error influences the reconstruction quality.

The translation of $V_3$ seems to correspond to a translation of $\alpha$.
In particular, considering the slope of $V_3$ in its linear part ($\frac{\Delta V_3}{\Delta\alpha}\approx0.037/\circ$),  the translation we applied ($\Delta V_3=-0.15$)  corresponds to  a new zero-lift angle of attack $\alpha_{zl}=4^\circ$.
\begin{figure}[H]
\setcounter{subfigure}{0}
    \centering
    \includegraphics[scale=0.25]{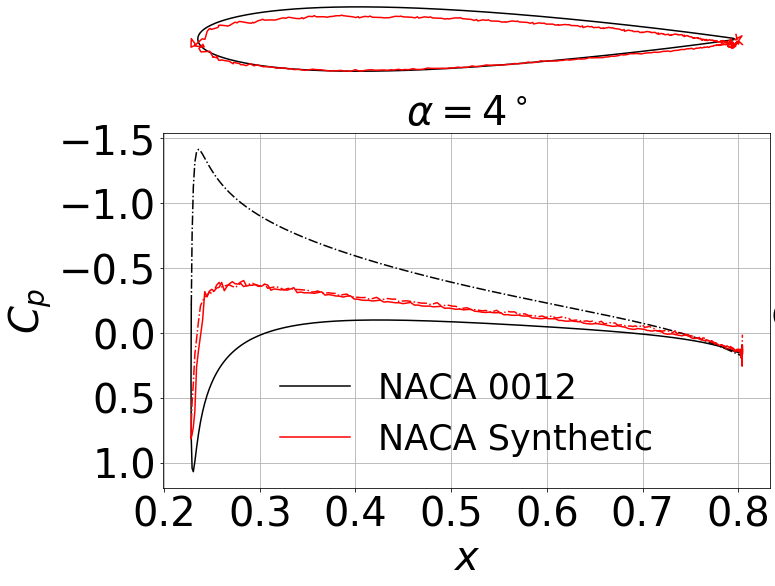}
    \caption{Output of the decoder with the translated variables at $\alpha=4^\circ$.}
    \label{Fig.latVarTra_4deg}
\end{figure}
Fig. \ref{Fig.latVarTra_4deg} shows that this is consistent with the results extracted  by the autoencoder for $\alpha=4^\circ$: the pressure coefficient  on the upper and lower surfaces of the synthetic airfoil  overlap, returning $C_l=0$.

The behaviour of $V_3$ resembles the lift curve, and the results corroborate this observation.

\subsection{Interpolation in the latent space}

It is not possible to  manipulate the latent space arbitrarily, and it is necessary to preserve the correlation between the latent variables. We developed  a  controlled strategy to extract new synthetic airfoils and flow-fields from the autoencoder.

\begin{figure}[H]
    \centering
    \subfloat[]{\includegraphics[scale=0.2]{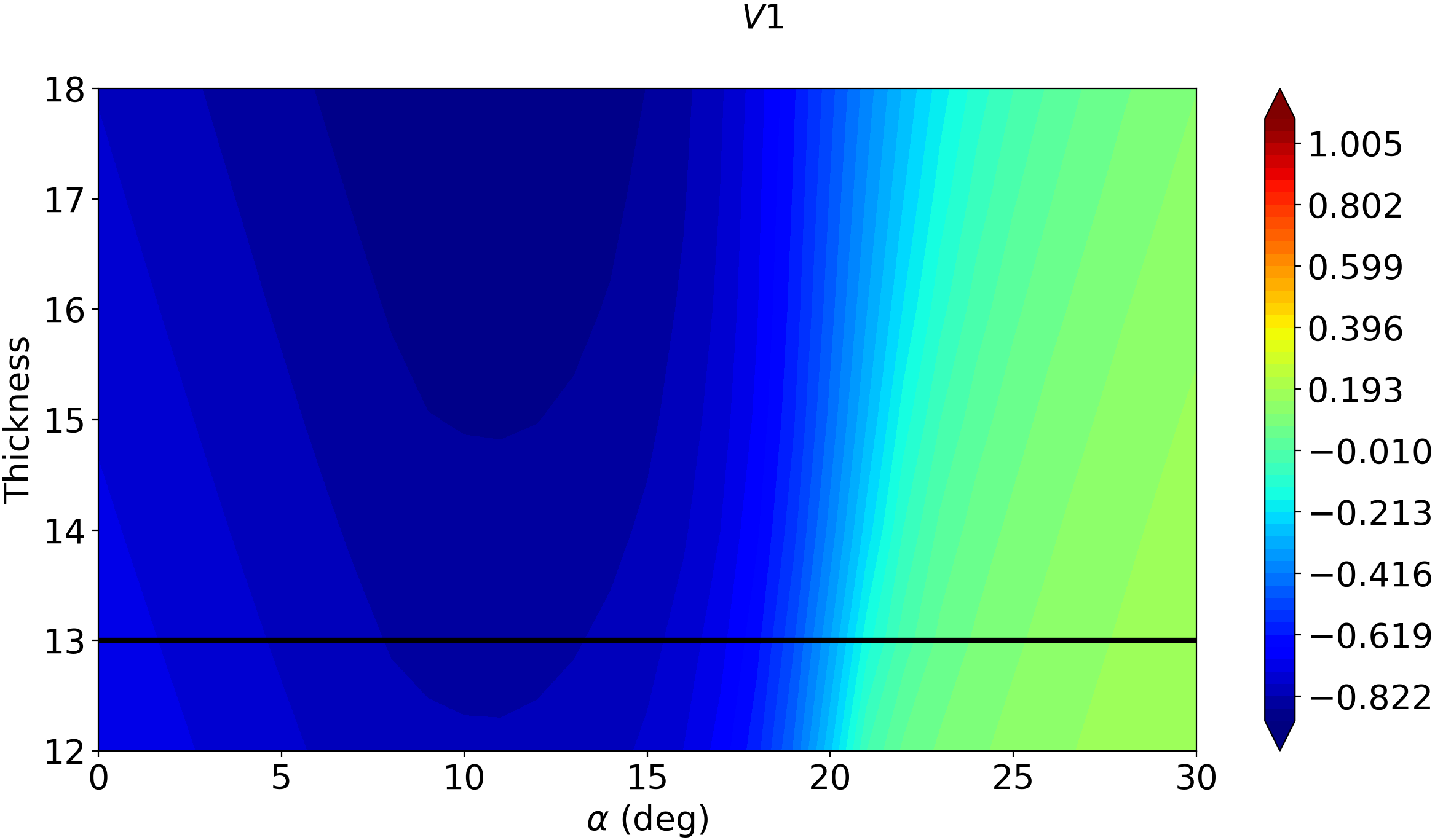}\label{subfig.V1_contour}}
    \subfloat[]{\includegraphics[scale=0.2]{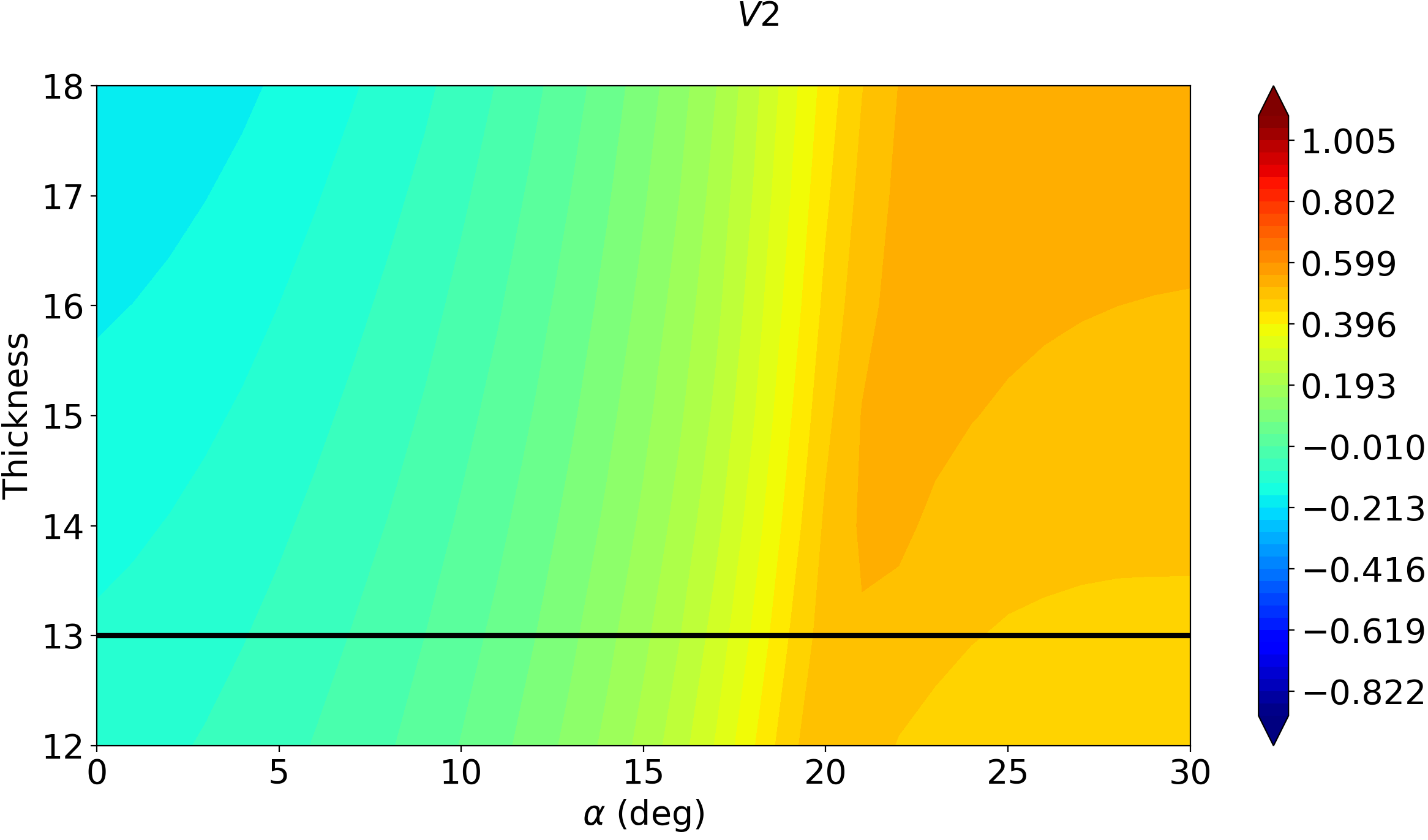}\label{subfig.V2_contour}}
    \subfloat[]{\includegraphics[scale=0.2]{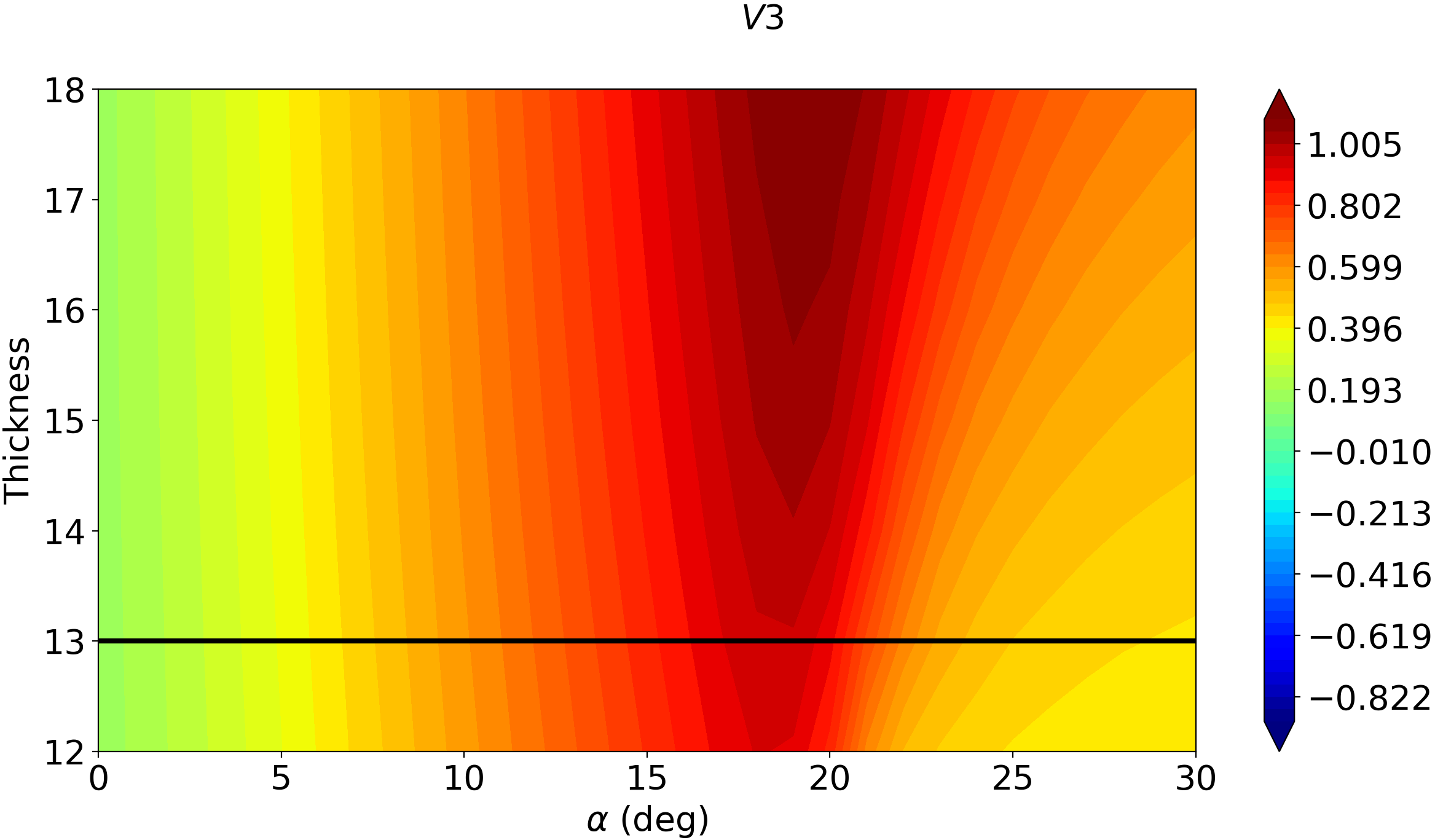}\label{subfig.V3_contour}}
    \caption{Contour mapping of the latent variables on the database parameters.}
    \label{fig:latVarInterp}
\end{figure}

\begin{figure}[H]
\setcounter{subfigure}{0}
    \centering
    \includegraphics[scale=0.6]{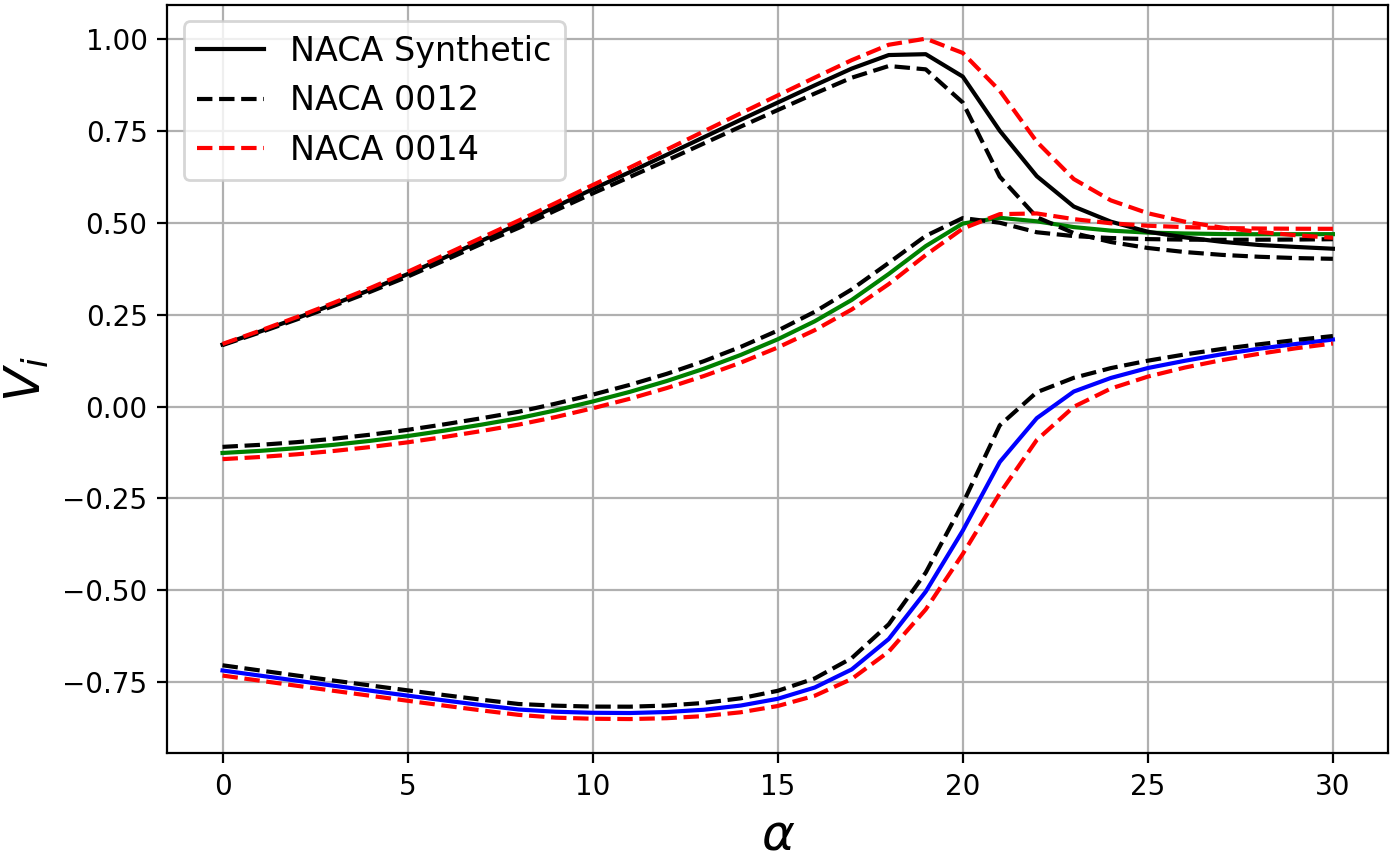}
    \caption{Interpolated latent variables at $t=13\%$ as function of the angle of attack $\alpha$. Comparison with the latent variables of the NACA 0012 and NACA 0014 airfoils. $V_1$: {\color{blue}---}; $V_2$: {\color{green}---}; $V_3$: --- .}
    \label{Fig.latVarInt_alpha}
\end{figure}

In the first step we created a mapping of the latent space on the database free parameters (airfoil thickness and angle of attack). The contour mapping is reported in Fig. \ref{fig:latVarInterp} for each latent variable of the latent space with $p=3$.
This mapping allows to easily interpolate, and we constructed the latent variables corresponding 
 to 13\% airfoil thickness (represented by the black line in Fig. \ref{fig:latVarInterp}).
The resulting latent variables as function of the angle of attack $\alpha$ are reported in Fig. \ref{Fig.latVarInt_alpha} and compared with the latent variables of the NACA 0012 (in dashed black lines) and NACA 0014 (in dashed red lines) airfoils.

\begin{figure}[H]
    \centering
    \includegraphics[scale=0.15]{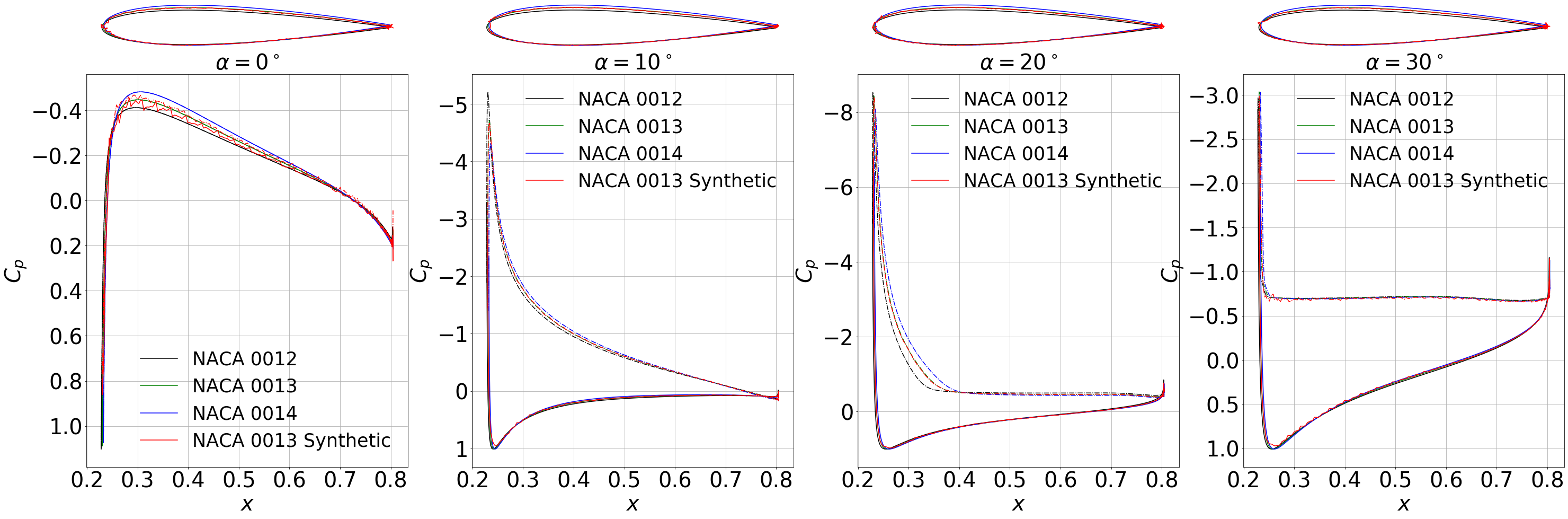}
    \caption{Comparison of the $C_p$ on the airfoil surfaces for $\alpha=0^\circ,\ 10^\circ,\ 20^\circ$ and $30^\circ$.}
    \label{fig:syn0013_cp}
\end{figure}

Starting from these interpolated latent variables the decoder generates solutions that can also be compared to the RANS solutions generated for the same {\it expected} profile (NACA 0013).
\begin{figure}[H]
    \centering
    \includegraphics[scale=0.15]{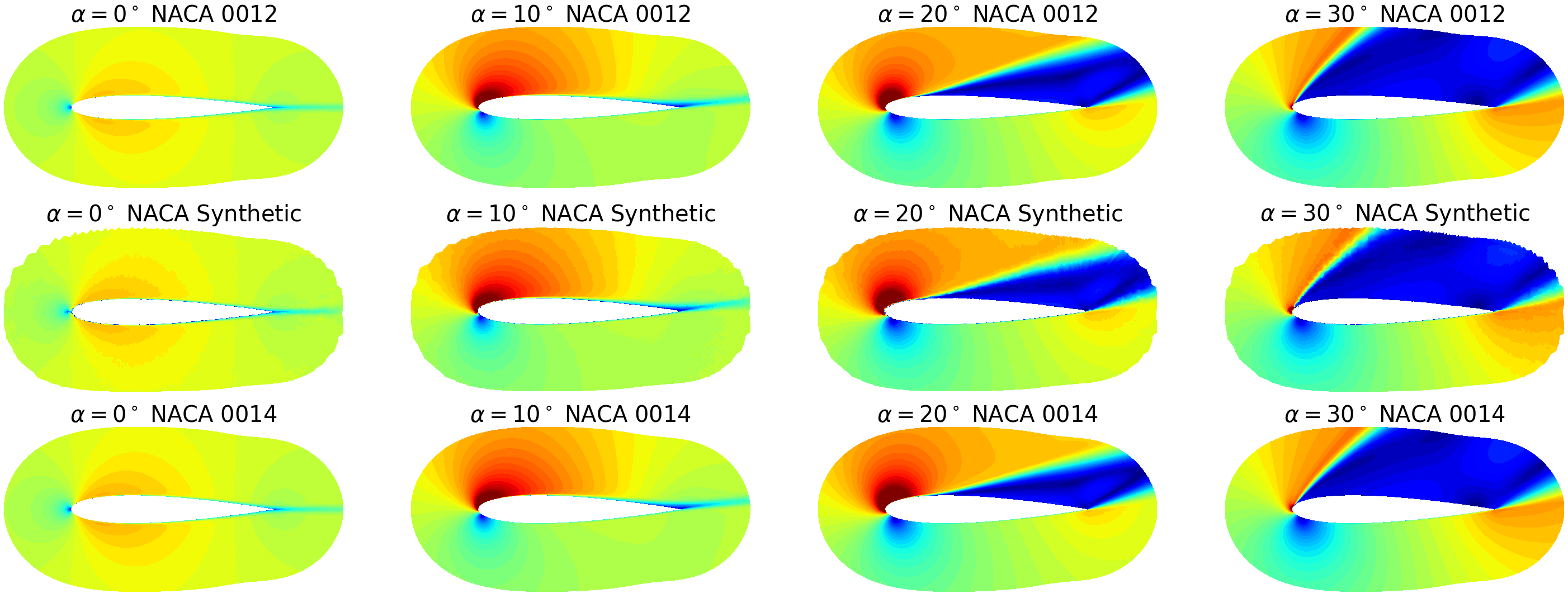}
    \caption{Mach number contour of the NACA 0012 and 0014 RANS solutions and NACA 0013 synthetic computed by the decoder for $\alpha=0^\circ,\ 10^\circ,\ 20^\circ$ and $30^\circ$.}
    \label{fig:res0013_mach}
\end{figure}
Fig. \ref{fig:res0013_mach} shows the Mach number contours for 4 angles of attack of the NACA 0012, 0014 and for the synthetic NACA airfoil generated by the decoder. In order to have a more quantitative analysis of the result, it is interesting to look at the pressure coefficient $C_p$ on the airfoil surface as reported in Fig. \ref{fig:syn0013_cp}.
The yellow curves refer to the NACA 0013 RANS solution, while the red curves refer to the decoder output (the synthetic NACA 0013), and they perfectly match  both in the linear regime (with the attached flow in the first two plots of Fig.\ref{fig:syn0013_cp}) and the non-linear regime and  stall conditions (the last two plots of Fig. \ref{fig:syn0013_cp}).
Also the geometry computed by the decoder is closely matching the expected NACA 0013 airfoil, and it is not changed with the angle of attack, showing that a coherent latent space manipulation leads to a consistent result.


\subsection{Extrapolation beyond the latent space}

\begin{figure}[H]
    \centering
    \subfloat[]{\includegraphics[scale=0.2]{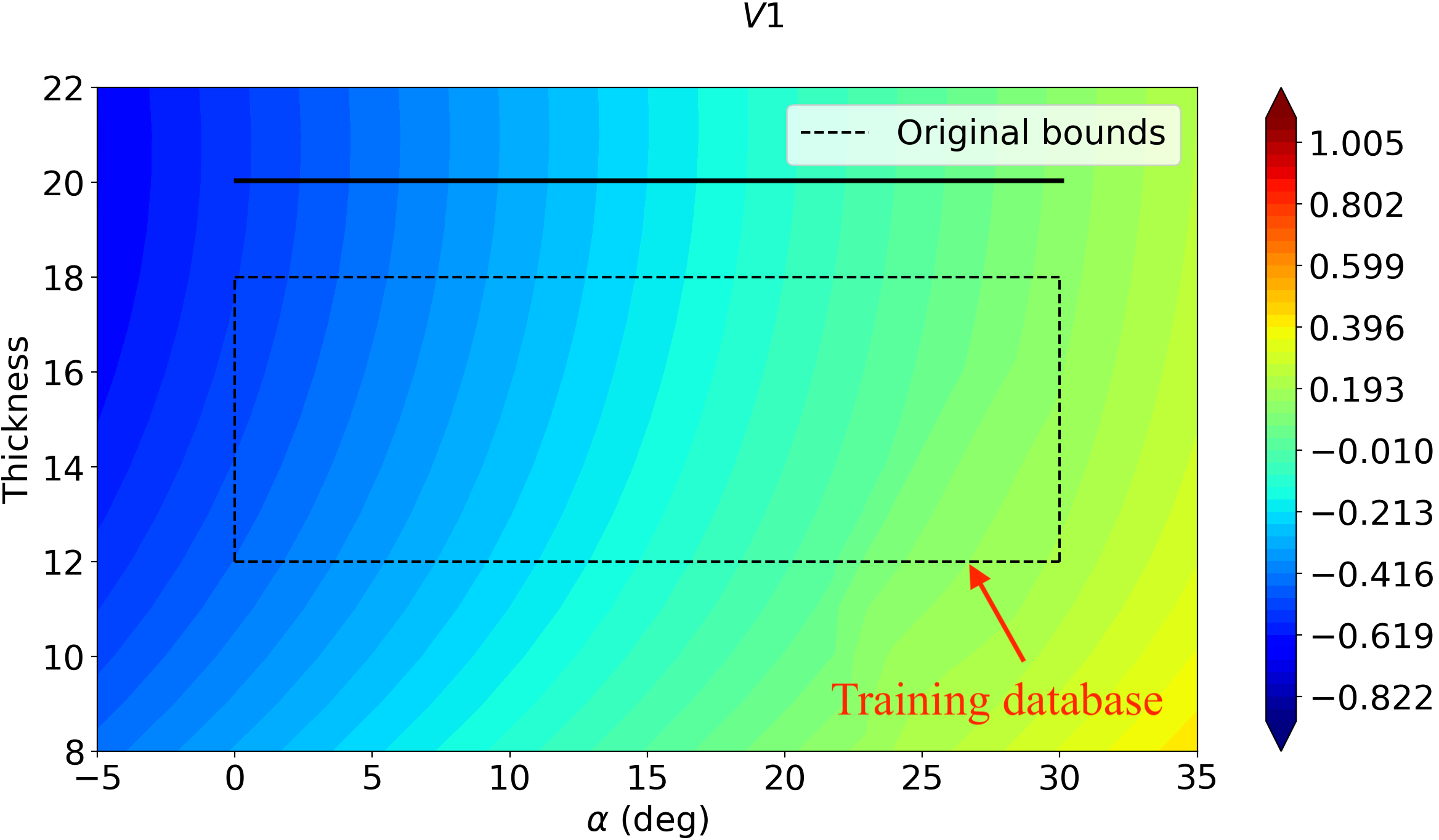}\label{subfig.V1_contour_extra}}
    \subfloat[]{\includegraphics[scale=0.2]{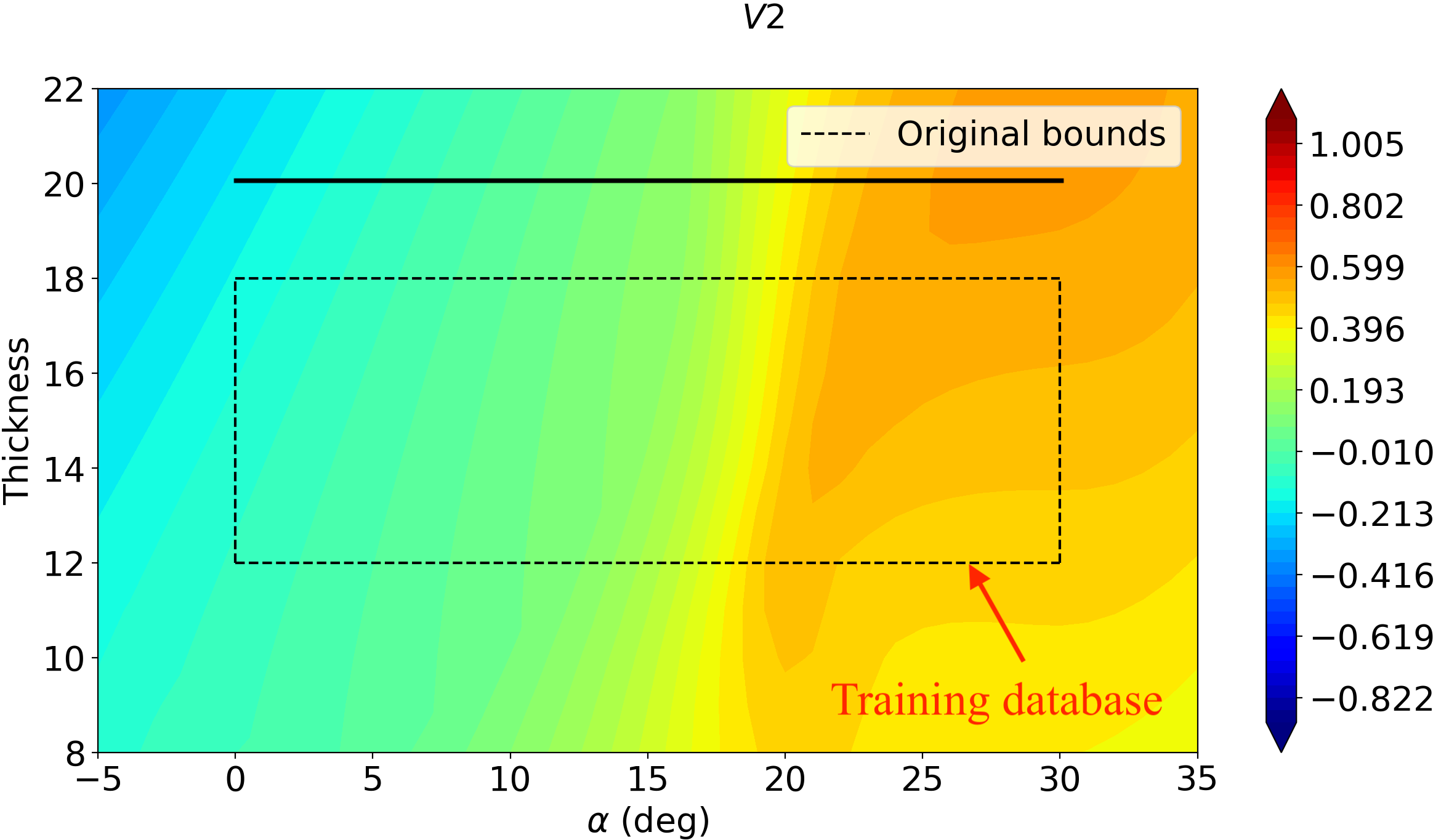}\label{subfig.V2_contour_extra}}
    \subfloat[]{\includegraphics[scale=0.2]{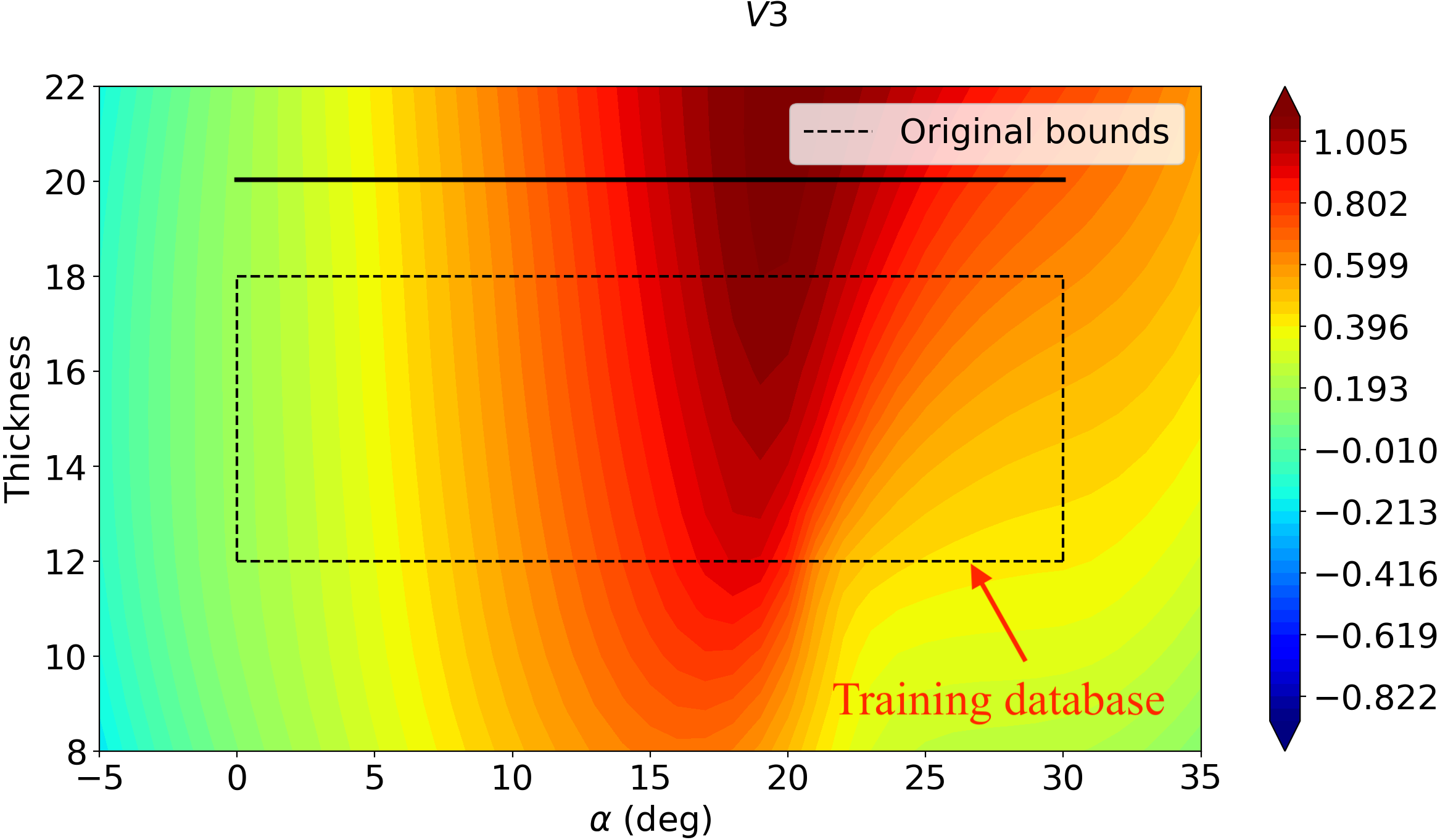}\label{subfig.V3_contour_extra}}
    
    \caption{Extrapolated contour mapping of the latent variables on the database parameters.}
    \label{fig:latVarExtrap}
\end{figure}

\begin{figure}[H]
    \centering
    \includegraphics[scale=0.3]{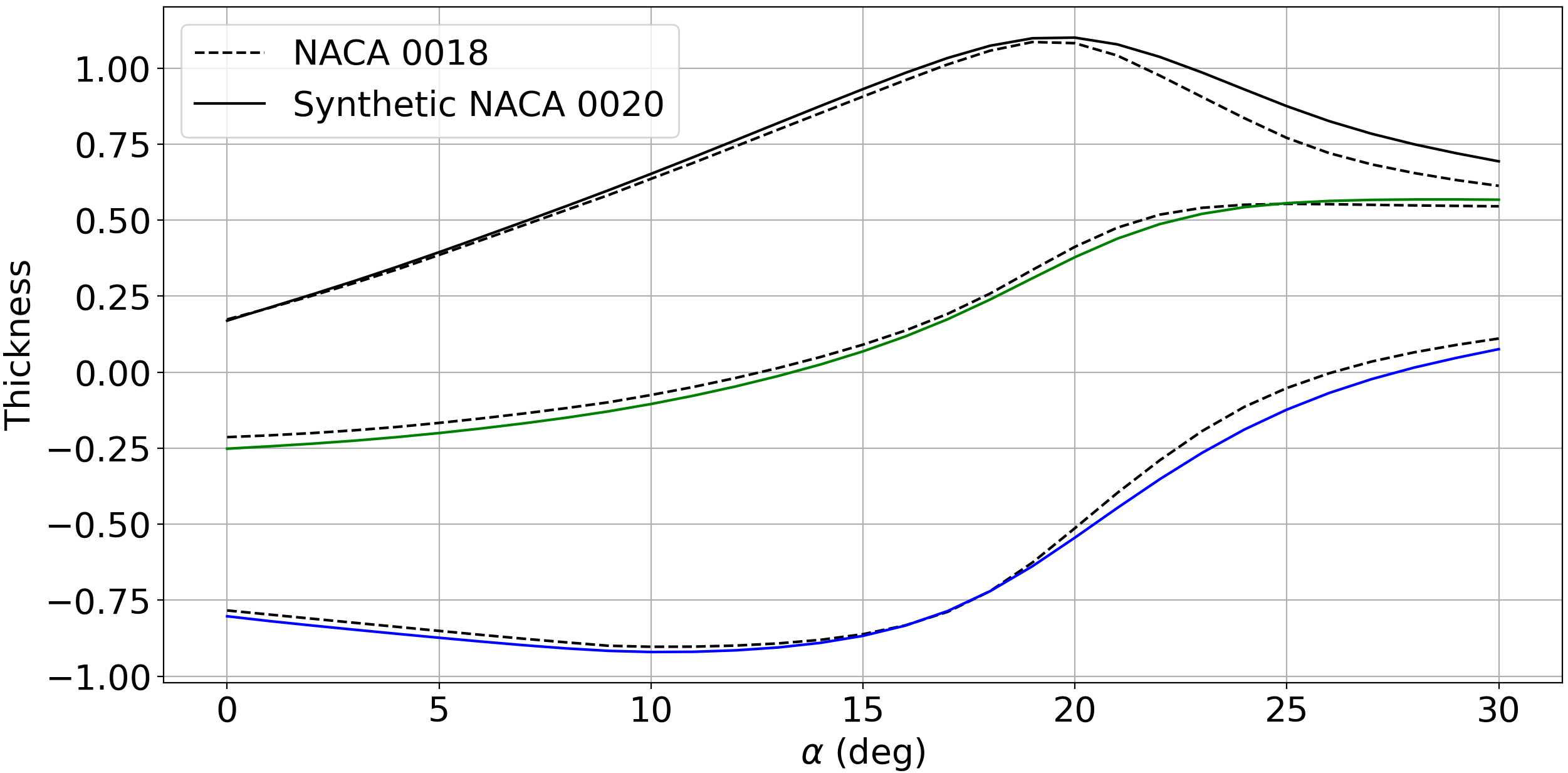}
    \caption{Extrapolated latent variables at $t=20\%$ as function of the angle of attack $\alpha$. Comparison with the latent variables of the NACA 0018 airfoil. $V_1$: {\color{blue}---}; $V_2$: {\color{green}---}; $V_3$: --- .}
    \label{fig:latVarExtrap_alpha}
\end{figure}

\begin{figure}[H]
    \centering
    \includegraphics[scale=0.15]{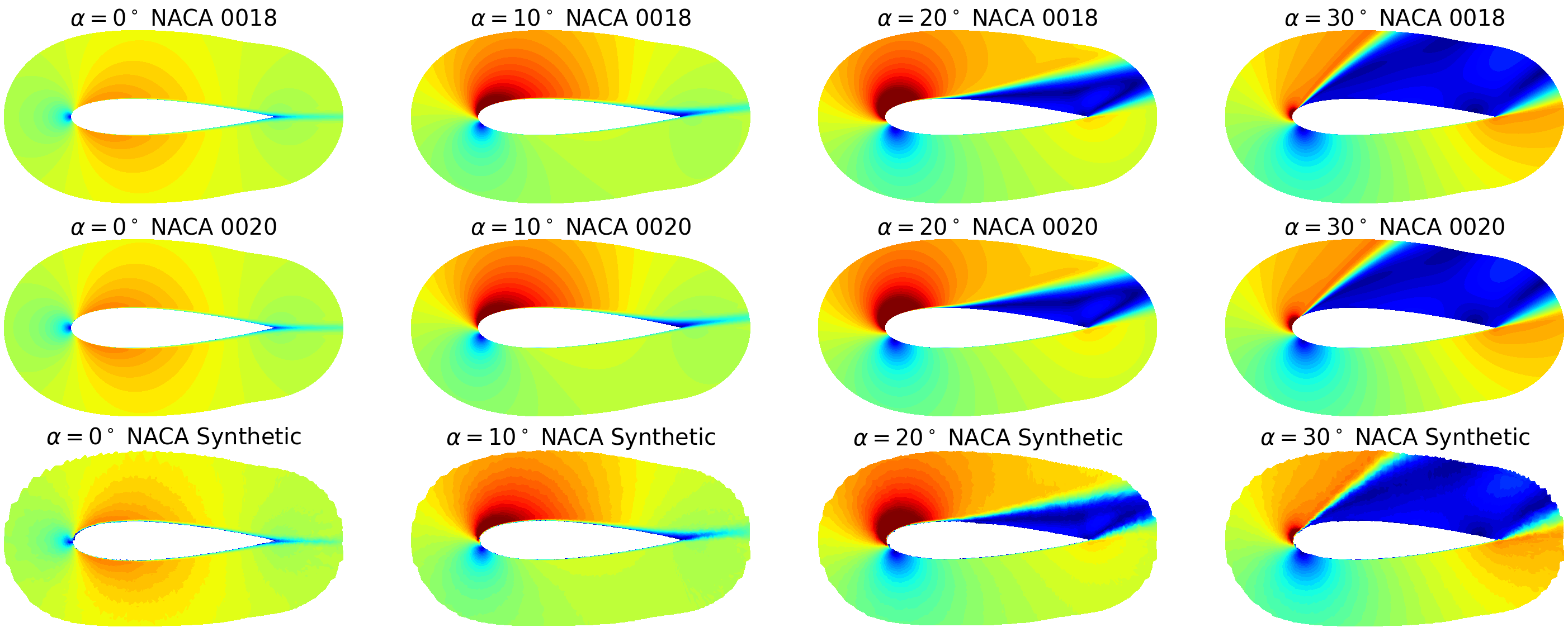}
    \caption{Comparison of the Mach number contour of the NACA 0018, NACA 0020 and NACA synthetic for $\alpha=0^\circ,\ 10^\circ,\ 20^\circ$ and $30^\circ$.}
    \label{fig:syn0020_mach}
\end{figure}

This result presented in the previous section clearly shows that it is possible to interpolate in the latent space learned by an autoencoder in order to obtain accurate aerodynamic predictions for a strongly non-linear phenomenon as the airfoil stall. But is it possible to extrapolate beyond the training dataset?  We performed a cubic extrapolation of the latent variables in the $\alpha$-$t$ plane, obtaining the new contour maps in Fig. \ref{fig:latVarExtrap}, where the black dashed rectangle represents the original boundaries of the input free parameters.
We extrapolated the values of the three latent variables at constant thickness $t=20\%$, represented by the continuous black line in Fig. \ref{fig:latVarExtrap}. These are plotted as function of $\alpha$ in Fig. \ref{fig:latVarExtrap_alpha}.
Fig. \ref{fig:syn0020_mach} shows the Mach number contours (for $\alpha=0^\circ,\ 10^\circ,\ 20^\circ$ and $30^\circ$) extracted from the RANS solutions of the NACA 0018 and 0020 and the ones generated by the decoder with the extrapolated latent variables in input.

\begin{figure}[H]
    \centering
    \includegraphics[scale=0.15]{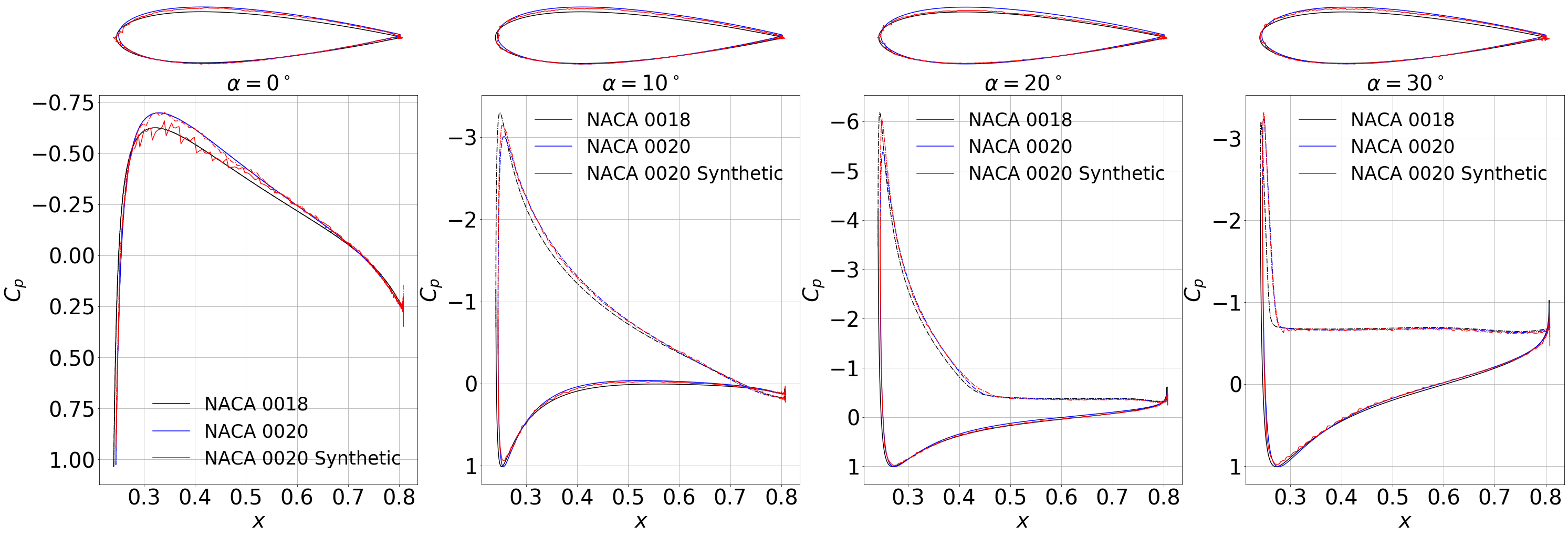}
    \caption{Comparison of the $C_p$ on the airfoil surfaces for $\alpha=0^\circ,\ 10^\circ,\ 20^\circ$ and $30^\circ$.}
    \label{fig:syn0020_cp}
\end{figure}

The contour of the Mach number of the synthetic flow-field appear to be in reasonable agreement with the RANS solution, however, in order to obtain a better evaluation of the results, we reported the $C_p$ on the airfoil surface in Fig. \ref{fig:syn0020_cp}.
Once again, the output of the decoder is in agreement with the RANS solution for the NACA 0020 for both the linear and non-linear regimes. In this extrapolatory case
 the accuracy is somewhat degraded with the respect to the interpolation case; specifically the maximum expansion region (the negative $c_p$ peak) is not exactly captured by the decoder.

\section*{Conclusions}
The results we presented in this work show that convolutional autoencoders are a powerful tool for the prediction of the aerodynamic performance of airfoils in both linear and non-linear aerodynamic regimes. We studied the sensitivity of the latent space of the trained autoencoder, showing that to pursue a physical interpretation it is necessary to use low-dimensional latent variables and it is preferable to use a database composed by raw data instead of images.

We showed that by using a randomized training dataset in which airfoil thickness and angle of attack vary, the autoencoder is able to automatically learn these parameters by organizing the latent space.

The autoencoder is also able to extract other physical information about the stall phenomenon: the latent variables are linear in the linear part of the phenomenon, and non-linear when stall occurs. Moreover, by using a latent space dimension $p=1$, the autoencoder learns that symmetric airfoils have the same behaviour for low angles of attack, and respond in a different way depending on their thickness towards the stall.

It is possible to interpolate and extrapolate in the latent space learned by the autoencoder in order to generate accurate aerodynamic predictions and flow-fields for synthetic airfoils  not seen in the training process. 

\bibliography{saetta_etal}

\begin{thebibliography}{17}
\newcommand{\enquote}[1]{``#1''}
\providecommand{\natexlab}[1]{#1}
\providecommand{\url}[1]{\texttt{#1}}
\providecommand{\urlprefix}{URL }
\expandafter\ifx\csname urlstyle\endcsname\relax
  \providecommand{\doi}[1]{\discretionary{}{}{}https://doi.org/#1}\else
  \providecommand{\doi}[1]{\discretionary{}{}{}\urlstyle{rm}\url{https://doi.org/#1}}\fi

\bibitem[{Duraisamy et~al.(2019)Duraisamy, Iaccarino, and Xiao}]{Duraisamy2019}
Duraisamy, K., Iaccarino, G., and Xiao, H., \enquote{Turbulence Modeling in the
  Age of Data,} \emph{Annual Review of Fluid Mechanics}, Vol.~51, 2019, pp.
  357--377.
\newblock \doi{10.1146/annurev-fluid-010518-040547}.

\bibitem[{Duraisamy et~al.(2015)Duraisamy, Zhang, and Singh}]{Duraisamy2015}
Duraisamy, K., Zhang, Z.~J., and Singh, A.~P., \enquote{New Approaches in
  Turbulence and Transition Modeling Using Data-driven Techniques,} \emph{AIAA
  SciTech}, 2015.
\newblock \doi{https://doi.org/10.2514/6.2015-1284}, 53rd AIAA Aerospace
  Sciences Meeting, 5-9 January 2015, Kissimmee, Florida.

\bibitem[{Tracey et~al.(2015)Tracey, Duraisamy, and Alonso}]{Tracey2015}
Tracey, B., Duraisamy, K., and Alonso, J.~J., \enquote{A Machine Learning
  Strategy to Assist Turbulence Model Development,} \emph{AIAA SciTech}, 2015.
\newblock \doi{10.2514/6.2015-1287}, 53rd AIAA Aerospace Sciences Meeting, 5-9
  January 2015, Kissimmee, Florida.

\bibitem[{Milano and Koumoutsakos(2002)}]{Milano2002}
Milano, M., and Koumoutsakos, P., \enquote{Neural Network Modeling for Near
  Wall Turbulent Flow,} \emph{Journal of Computational Physics}, Vol. 182,
  2002, pp. 1--26.
\newblock \doi{10.1006/jcph.2002.7146}.

\bibitem[{Yan et~al.(2019)Yan, Zhu, Kuang, and Wang}]{Yan2019}
Yan, X., Zhu, J., Kuang, M., and Wang, X., \enquote{Aerodynamic shape
  optimization using a novel optimizer based on machine learning techniques,}
  \emph{Aerospace Science and Technology}, Vol.~86, 2019, pp. 826--835.
\newblock \doi{https://doi.org/10.1016/j.ast.2019.02.003}.

\bibitem[{Li et~al.(2020)Li, Zhang, Martins, and Shu}]{Li2020}
Li, J., Zhang, M., Martins, J. R. R.~A., and Shu, C., \enquote{Efficient
  Aerodynamic Shape Optimization with Deep-learning-based Geometric Filtering,}
  \emph{AIAA Journal}, Vol.~58, No.~10, 2020.
\newblock \doi{10.2514/1.J059254}.

\bibitem[{Noack(2019)}]{Noack2018}
Noack, B.~R., \enquote{{Closed-Loop Turbulence Control-From Human to Machine
  Learning (and Retour)},} \emph{{Proceedings of the 4th Symposium on Fluid
  Structure-Sound Interactions and Control (FSSIC)}}, edited by Zhou, Y.,
  Kimura, M., Peng, G.~Lucey, A.D., .~Huang, and L., {Springer}, 2019, pp.
  23--32.
\newblock \urlprefix\url{https://hal.archives-ouvertes.fr/hal-02398734}.

\bibitem[{Verma et~al.(2018)Verma, Novati, and Koumoutsakos}]{Verma2018}
Verma, S., Novati, G., and Koumoutsakos, P., \enquote{Efficient collective
  swimming by harnessing vortices through deep reinforcement learning,}
  \emph{Proceedings of the National Academy of Sciences}, Vol. 115, No.~23,
  2018, pp. 5849--5854.
\newblock \doi{10.1073/pnas.1800923115},
  \urlprefix\url{https://www.pnas.org/doi/abs/10.1073/pnas.1800923115}.

\bibitem[{Matalanis et~al.(2016)Matalanis, Bowles, Jee, Min, Kuczek, Croteau,
  Wake, Crittenden, Glezer, and Lorber}]{matalanis2016}
Matalanis, C.~G., Bowles, P.~O., Jee, S., Min, B.-Y., Kuczek, A.~E., Croteau,
  P.~F., Wake, B.~E., Crittenden, T., Glezer, A., and Lorber, P.~F.,
  \enquote{Dynamic Stall Suppression Using Combustion-Powered Actuation
  (COMPACT),} Tech. rep., 2016.

\bibitem[{Economon et~al.(2016)Economon, Palacios, Copeland, Lukaczyk, and
  Alonso}]{Economon2016SU2}
Economon, T.~D., Palacios, F., Copeland, S.~R., Lukaczyk, T.~W., and Alonso,
  J.~J., \enquote{SU2: An Open-Source Suite for Multiphysics Simulation and
  Design,} \emph{AIAA Journal}, Vol.~54, No.~3, 2016, pp. 828--846.
\newblock \doi{10.2514/1.J053813}.

\bibitem[{Rumsey et~al.(2019)Rumsey, Slotnick, and Sclafani}]{Rumsey2019}
Rumsey, C.~L., Slotnick, J.~P., and Sclafani, A.~J., \enquote{Overview and
  Summary of the Third AIAA High Lift Prediction Workshop,} \emph{Journal of
  Aircraft}, Vol.~56, No.~2, 2019, pp. 621--644.
\newblock \doi{10.2514/1.C034940},
  \urlprefix\url{https://doi.org/10.2514/1.C034940}.

\bibitem[{Lee and Carlberg(2020)}]{kookjin2020}
Lee, K., and Carlberg, K.~T., \enquote{Model reduction of dynamical systems on
  nonlinear manifolds using deep convolutional autoencoders,} \emph{Journal of
  Computational Physics}, Vol. 404, 2020, p. 108973.
\newblock \doi{https://doi.org/10.1016/j.jcp.2019.108973},
  \urlprefix\url{https://www.sciencedirect.com/science/article/pii/S0021999119306783}.

\bibitem[{Bhatnagar et~al.(2019)Bhatnagar, Afshar, Pan, Duraisamy, and
  Kaushik}]{Bhatnagar2019}
Bhatnagar, S., Afshar, Y., Pan, S., Duraisamy, K., and Kaushik, S.,
  \enquote{Prediction of aerodynamic flow fields using convolutional neural
  networks,} \emph{Computational Mechanics}, Vol.~64, No.~2, 2019, pp.
  525--545.
\newblock \doi{10.1007/s00466-019-01740-0},
  \urlprefix\url{https://doi.org/10.1007/s00466-019-01740-0}.

\bibitem[{Tangsali et~al.(2020)Tangsali, Krishnamurthy, and
  Hasnain}]{Tangsali2020}
Tangsali, K., Krishnamurthy, V.~R., and Hasnain, Z., \enquote{{Generalizability
  of Convolutional Encoder–Decoder Networks for Aerodynamic Flow-Field
  Prediction Across Geometric and Physical-Fluidic Variations},} \emph{Journal
  of Mechanical Design}, Vol. 143, No.~5, 2020.
\newblock \doi{10.1115/1.4048221},
  \urlprefix\url{https://doi.org/10.1115/1.4048221}, 051704.

\bibitem[{Agostini(2020)}]{Agostini2020}
Agostini, L., \enquote{Exploration and prediction of fluid dynamical systems
  using auto-encoder technology,} \emph{Physics of Fluids}, Vol.~32, No.~6,
  2020, p. 067103.
\newblock \doi{10.1063/5.0012906},
  \urlprefix\url{https://doi.org/10.1063/5.0012906}.

\bibitem[{Paszke et~al.(2019)Paszke, Gross, Massa, Lerer, Bradbury, Chanan,
  Killeen, Lin, Gimelshein, Antiga, Desmaison, Kopf, Yang, DeVito, Raison,
  Tejani, Chilamkurthy, Steiner, Fang, Bai, and Chintala}]{Pytorch2019}
Paszke, A., Gross, S., Massa, F., Lerer, A., Bradbury, J., Chanan, G., Killeen,
  T., Lin, Z., Gimelshein, N., Antiga, L., Desmaison, A., Kopf, A., Yang, E.,
  DeVito, Z., Raison, M., Tejani, A., Chilamkurthy, S., Steiner, B., Fang, L.,
  Bai, J., and Chintala, S., \enquote{PyTorch: An Imperative Style,
  High-Performance Deep Learning Library,} \emph{Advances in Neural Information
  Processing Systems 32}, Curran Associates, Inc., 2019, pp. 8024--8035.
\newblock
  \urlprefix\url{http://papers.neurips.cc/paper/9015-pytorch-an-imperative-style-high-performance-deep-learning-library.pdf}.

\bibitem[{Newson et~al.(2020)Newson, Almansa, Gousseau, and
  Ladjal}]{Newson2020}
Newson, A., Almansa, A., Gousseau, Y., and Ladjal, S., \enquote{Processing
  Simple Geometric Attributes with Autoencoders,} \emph{Journal of Mathematical
  Imaging and Vision}, Vol.~62, No.~3, 2020, pp. 293--312.
\newblock \doi{10.1007/s10851-019-00924-w},
  \urlprefix\url{https://doi.org/10.1007/s10851-019-00924-w}.

\end{thebibliography}

\end{document}